\documentclass[aps,prd,twocolumn,floatfix,nofootinbib,superscriptaddress,tightenlines]{revtex4}

\usepackage[dvipsnames]{xcolor}
\usepackage{amsmath}
\usepackage{dcolumn}
\usepackage{lipsum}
\usepackage{amssymb}
\usepackage{url}
\usepackage{epsfig}
\usepackage{graphicx}
\usepackage{amsmath}
\usepackage{bm}
\usepackage{setspace}
\usepackage{lscape}
\usepackage{amsthm}
\usepackage{bbold}
\usepackage{dcolumn}
\usepackage{epsfig}
\usepackage{graphics}
\usepackage{graphicx}
\usepackage{longtable}
\usepackage{svg}

\usepackage{bm}
\usepackage{xspace}
\usepackage{cancel}
\usepackage{float}
\usepackage{multirow}
\definecolor{darkgreen}{rgb}{0,0.5,0}
\definecolor{purple}{rgb}{0.5,0,0.5}
\definecolor{nblue}{rgb}{0.0,0.0,0.50}
\definecolor{scarlet}{rgb}{1.0,0.2,0}
\definecolor{darkmagenta}{rgb}{0.55, 0.0, 0.55}
\definecolor{darkolivegreen}{rgb}{0.33, 0.42, 0.18}
\definecolor{darkcandyapplered}{rgb}{0.64, 0.0, 0.0}
\usepackage[colorlinks=true, pdfstartview=FitV, linkcolor=purple, citecolor= purple, urlcolor=blue]{hyperref}

\def\beq{\begin{equation}} \def\eeq{\end{equation}}
\def\beqn{\begin{eqnarray}} \def\eeqn{\end{eqnarray}}

\def\beq{\begin{equation}} 
\def\eeq{\end{equation}} 
\def\beqn{\begin{eqnarray}} 
\def\eeqn{\end{eqnarray}} 
 
\def\to{\rightarrow}
\def\nn{\nonumber}

\begin{document}
\title{Constraining fragmentation functions through hadron-photon production at higher-orders}

\author{Salvador A. Ochoa-Oregon}
\email{salvadorochoa.fcfm@ms.uas.edu.mx}
\affiliation{Facultad de Ciencias F\'isico-Matem\'aticas, Universidad Aut\'onoma de Sinaloa, Ciudad Universitaria, Culiac\'an, Sinaloa 80000,
M\'exico.}

\author{David F. Renter\'ia-Estrada}
\email{davidrenteria.fcfm@uas.edu.mx}
\affiliation{Facultad de Ciencias F\'isico-Matem\'aticas, Universidad Aut\'onoma de Sinaloa, Ciudad Universitaria, Culiac\'an, Sinaloa 80000,
M\'exico.} 

\author{Roger J. Hern\'andez-Pinto}
\email{roger@uas.edu.mx}
\affiliation{Facultad de Ciencias F\'isico-Matem\'aticas, Universidad Aut\'onoma de Sinaloa, Ciudad Universitaria, Culiac\'an, Sinaloa 80000,
M\'exico.}

\author{German F. R. Sborlini}
\email{german.sborlini@usal.es}
\affiliation{Departamento de F\'isica Fundamental e IUFFyM, Universidad de Salamanca, 
37008 Salamanca, Spain.}
\affiliation{Escuela de Ciencias, Ingenier\'ia y Diseño, Universidad Europea de Valencia, \\ Paseo de la Alameda 7, 46010 Valencia, Spain.}

\begin{abstract}
In certain situations, such as one-particle inclusive processes, it is possible to model the hadronization through Fragmentation Functions (FFs), which are universal non-perturbative functions extracted from experimental data through advanced fitting techniques. Constraining the parameters of such fits is crucial to reduce the uncertainties, and provide reliable and accurate FFs. In this article, we explore strategies to relate pion and FFs for other hadrons (in particular, kaons), comparing cross-section ratios imposing proper kinematical cuts. We exploit the phenomenology of photon-hadron production at colliders, including up to NLO QCD and LO QED corrections, and make use of accurate formulae to reconstruct the partonic momentum fractions. By studying different cuts, we manage to isolate the contribution of $u$-started FFs. Then, we relate the ratios of the $z$-spectrum for pion and kaon production, with the corresponding FFs ratios. The methodology described in this article can be used to relate FFs for any pair of hadrons, and could be further explored to keep track of the flavour of the partons undergoing the hadronization. 

\end{abstract}

\maketitle               


\section{Introduction and motivation}
\label{sec:introduction}
Nowadays, one of the greatest challenges of humankind is related to a proper understanding of the microscopic behaviour of matter. During the last five decades, a great knowledge was acquired through various generations of precise experiments \cite{Gianotti:2002xx,ILC:2007bjz,Roloff:2018dqu,CEPCStudyGroup:2018ghi,FCC:2018byv,FCC:2018evy,FCC:2018vvp,FCC:2018bvk,Strategy:2019vxc}, as well as an extensive development of theoretical models. One of these theories, the Standard Model (SM), is currently considered as one of the most precise descriptions of Nature. However, there are several aspects of SM that escape our technical skills.

One of these aspects is related to the extraction of exact solutions. Since it is a quantum field theory (QFT), with a non-Abelian gauge group, the underlying mathematical structure is rather complex. For this reason, several approximated strategies were deployed, including perturbation theory. This method is particularly useful to describe high-energy collisions of particles, although unable to provide a reliable quantification of the low-energy interactions involved in the hadronization process. This problem is partially by-passed using the so-called parton model \cite{Drell:1969ca,Feynman:1969wa,Feynman:1970fm,Altarelli:1977zs}, which proposed to describe hadronic cross-sections in terms of convolutions of \emph{partonic cross-section} (perturbative object describing the scattering of fundamental particles) and distribution functions (non-perturbative components that indicate the distribution of fundamental particles within hadrons). 

The non-perturbative distribution functions can be of two different types. In one side, we have the Parton Distribution Functions (PDFs), $f^h_a(x,\mu)$, which roughly quantify the probability distribution of extracting a parton of flavour $a$ from a hadron $h$ with momentum fraction $x$ at the energy scale $\mu$. On the other side, there are Fragmentation Functions (FFs), $d^h_a(z,\mu)$, that indicate the probability distribution of originating a hadron $h$ with momentum fraction $z$ through the hadronization of a parton $a$ at the energy scale $\mu$. The purpose of this article is to provide novel strategies for the precise determination of the latest.

Highly-accurate determination of the non-perturbative distributions is not a simple task but they are crucial to understand high energetic collider experiments. Since they cannot be fully predicted from first principles, their extraction need to consider theoretical and experimental inputs. From the experimental community, different experiments can provide the data for different mesons coming from Semi-Inclusive Deep Inelastic Scattering (SIDIS) \cite{HERMES:2012uyd,COMPASS:2016xvm,COMPASS:2016crr} and hadron-hadron collisions \cite{STAR:2013zyt,STAR:2011iap,STAR:2009vxb,STAR:2006xud,ALICE:2012wos,ALICE:2014juv,PHENIX:2007kqm}, in the case of PDFs, and it is important to include Single-Inclusive Annihilation (SIA) \cite{OPAL:1994zan,TASSO:1988jma,BaBar:2013yrg,Belle:2013lfg,ALEPH:1994cbg,DELPHI:1998cgx,SLD:1998coh,OPAL:1999zfe,TPCTwoGamma:1986evp,TPCTwoGamma:1988yjh} results for a global QCD analysis of the FFs. On the other side, theoretical predictions at LO, NLO \cite{Hirai:2007cx,Albino:2005mv,Albino:2005me,Kniehl:2000fe,Kretzer:2000yf,deFlorian:2014xna,deFlorian:2017lwf} are available, and efforts in order to include up to NNLO corrections \cite{Borsa:2022vvp,AbdulKhalek:2022laj,Ritzmann:2014mka} are under investigation. Matching between theory and experiment assumes a particular modelling of the non-perturbative objects. There are different ansatzes to characterize them and nowadays there are also proposals with no ansatz such as those using Neural Network frameworks \cite{Bertone:2017bme}. Although, all groups provide a good agreement to describe the experimental data, among error bands, there is still the question about the correct theoretical description of the PDFs and FFs \footnote{A further review can be found in Ref. \cite{Metz:2016swz}.}. Furthermore, it is known that the inclusion of data from experiments at different energies (for instance, RHIC at 500 GeV and LHC at 13 TeV) could lead to incompatibilities of the models and break the expected naive universality of the FFs. When the processes included in the experiment-theory confrontation are inclusive enough, the discrepancies can be reduced by manipulating the factorization scales \cite{Borsa:2021ran}. The situation is expected to worsen when less inclusive processes are considered, which posses a clear challenge for understanding the hadronization and forces to explore new techniques to describe this phenomena.
    
With this panorama in mind, the purpose of the present article consists in exploring strategies for imposing stricter constraints on FFs. In particular, it is important to recall that there are experimental efforts to determine \emph{FFs per flavour} (i.e. keeping track of the flavour of the parton undergoing the hadronization). Since flavour-tagged SIA data restrict heavy quark FFs, it is worth investigating phenomenological disentanglement of FFs, at least for the leading flavour. Therefore, we exploit the fact that pion FFs are determined with a relatively small error w.r.t. fragmentation processes involving heavier hadrons. Also, that the presence of hard photons in the final state could be used as a clear probe to reveal information of the hard scattering process (i.e. parton level collisions). For these reasons, we use previous results of $\gamma + \pi$ production at colliders \cite{Binoth:2002wa,Arleo:2006xb}, including up to NLO QCD + LO QED corrections \cite{deFlorian:2010vy,Renteria-Estrada:2021rqp}. These calculations are implemented into a semi-automatic code that allows the user to change the FFs set, thus switching to $\gamma + h$ with a generic hadron $h$ in the final state. In this paper, we specify the case $h=K$ since it is the next-to-lightest meson. Then, we make use of reconstruction formula to rewrite the momentum fraction of the parton undergoing the fragmentation, i.e. $z$, in terms of experimentally-accessible quantities. In concrete, we take advantage of our previous studies in Refs. \cite{deFlorian:2010vy,Renteria-Estrada:2022scipost} to achieve a reliable quantification of this momentum fraction.

Once the differential distributions for $\gamma + \pi$ and $\gamma + K$ are calculated in terms of the reconstructed momentum fraction $z$, we go one step further and quantify the impact of kinematical cuts to discriminate $q$ and $g$ initiated sub-processes. With this, we aim to relate the ratio of the cross-sections with ratios of FFs involving specific partons undergoing the hadronization. Similar analysis using photon-jet or hadron-jet correlations were previously performed to extract information about FFs \cite{Belghobsi:2009hx,Arleo:2013tya,Klasen:2014xfa}. The spirit of the exploration proposed in this paper is motivated by very recent results on jet structure analysis and discrimination of gluon/quark initiated jets exploiting machine-learning (ML) techniques \cite{Ying:2022jvy,Cheung:2022dil,Yang:2022yfr,Bright-Thonney:2022xkx}.

After this introduction, we can summarize the three main ideas of this paper:
\begin{enumerate}
    \item Requiring the presence of an isolated photon in the process $p+p \to \gamma + h$, we can accurately reconstruct the partonic momentum fractions $x$ and $z$ in terms of experimentally-accessible variables.
    \item The pion FFs are very well determined, in comparison to fragmentation into heavier hadrons (such as $K$ mesons).
    \item Imposing proper kinematical cuts, which could involve restricting the values of $x$ and/or $z$, we can factorize the dependence of certain FFs ($d^h_u$, $d^g_a$, etc.) at the differential cross-section level.
\end{enumerate}
Then, the main motivation of this paper is to find relations between parton-to-pion and parton-to-hadron FFs from the study of ratios of the $z$-spectra of $p+p \to \gamma + \pi$ and $p+p \to \gamma + h$.

The outline of this paper is the following. In Sec. \ref{sec:Analysis}, we carefully describe the details of the implementation. Also, we explain the different cuts implemented and their relevance in disentangling the production channels. Then, in Sec. \ref{sec:FFconstraints}, we compare the ratio of the distributions in the reconstructed momentum fraction $z$ against ratios of FFs. We discuss the similarities among these plots and suggest potential improvements in constraining the FFs from the cross-section shape. In particular, in Sec. \ref{ssec:FFimproved}, we use the reconstructed momentum fractions discussed in Refs. \cite{deFlorian:2010vy,Renteria-Estrada:2022scipost} to impose specific cuts and extract information about FFs ratios. Finally, in Sec. \ref{sec:conclusions}, we present the conclusions of this work and depict possible future research strategies to improve the quality of kaon (an other heavier hadrons) fragmentation functions.


\section{Computational details and phenomenological analysis}
\label{sec:Analysis}
As a first step, it is necessary to calculate the hadron+photon fully-differential cross-section. For this purpose, we use the factorization theorem \cite{Collins:1985ue} and compute the hadronic cross-section in terms of convolutions of partonic cross-sections, PDFs and FFs. In this work, we rely on the implementation presented in Refs. \cite{deFlorian:2010vy,Renteria-Estrada:2021rqp}, centering our attention on the production of prompt photons. This code incorporates NLO QCD and LO QED corrections, which turned out to be relevant to provide a reliable phenomenological description in the high-energy regime \cite{Renteria-Estrada:2021rqp}. The requirement of having a prompt or hard photon in the final state is crucial for accessing the parton-level kinematics, and hence reconstructing with high-precision the parton momentum fractions in terms of the momenta of the photon and the produced hadron. In concrete, we perform the cross-section simulation of the processes
\begin{eqnarray}
p+p&\to&\gamma+\pi\,,\\
p+p&\to&\gamma+K\,,
\end{eqnarray}
with $\pi = \{\pi^\pm,\,\pi^0\}$ and $K=\{K^\pm,\,K^0\}$. To restrict the contamination due to events in which the photon comes from a hadronic decay, we implemented the smooth-cone isolation algorithm \cite{Frixione:1998jh}. In particular, we used
\begin{eqnarray}
    \xi(r)=\epsilon_rE_T^{\gamma}\left(\frac{1-\cos{r}}{1-\cos{r_0}}\right)^4\, ,
\end{eqnarray}
as cut function to restrict the amount of hadronic energy surrounding the photon. This cut depends on transverse energy photon $E_T^\gamma$. Additionally, we set the parameters $\epsilon_r=1$ and $r_0=0.4$, following the choice done in previous analysis of this process. Likewise, 
we define  
\begin{eqnarray}
    \mu = \frac{p_T^\gamma+p_T^h}{2}\,,
    \label{eq:MUdef}
\end{eqnarray}
as the typical energy scale of the hard process, i.e. we average the transverse momenta of the hadron and the photon ($h=\pi,\,K$). So, the default configuration is given by $\mu=\mu_F=\mu_R=\mu_I$. 

Regarding the hadronic center-of-mass (c.m.) energy, we used $E_{CM}=\sqrt{S_{CM}}=13$ TeV, as well as the default kinematic cuts,
\begin{eqnarray}
&& \{|\eta^h|,|\eta^\gamma|\} \leq 2.5 \, ,  
\\ && 2 \, {\rm GeV} \geq p_T^h \geq 15 \, {\rm GeV} \, ,
\\ && p_T^\gamma \geq 30 \, {\rm GeV} \, ,
\label{eq:CUTSexperimentales2}
\end{eqnarray}
which corresponds to the range covered by LHC Run II. With respect to azimuthal angles, we have imposed the restriction $\Delta\phi=|\phi^h-\phi^\gamma|\geq 2$, with the purpose of keeping events which are close to a back-to-back configuration. 

The non-perturbative effects due to the low-energy interactions driving hadron internal structure were captured inside the definition of PDFs and FFs. These functions were implemented using the unified framework {\tt LHAPDF} \cite{Buckley:2014ana,Andersen:2014efa}. In our simulations, we used two different PDF sets: {\tt NNPDF40\_lo\_as\_01180} \cite{Kassabov:2022pps} for LO QCD, whilst we relied on {\tt NNPDF31\_nlo\_as\_0118\_luxqed} \cite{Campbell:2018wfu,Bertone:2017bme,Manohar:2017eqh} PDF set when including LO QED + NLO QCD corrections. The impact of the hadronization of hard partons into light mesons (pions, kaons, etc.) was quantified through the FFs of the DSS collaboration: we relied on {\tt DSS2014} \cite{deFlorian:2014xna} and {\tt DSS2017} \cite{deFlorian:2017lwf} to describe pion and kaon production, respectively. In any case, we would like to emphasize that the codes used allow to implement any PDFs or FFs set.


\subsection{Differential cross-section vs. partonic momentum fraction \emph{z}}
\label{ssec:DistributionsZ}
Since the aim of this work is to impose constraints on FFs, the first study consists in the examination of the $z$-spectrum of the hadronic cross-sections. Instead of the Monte-Carlo (MC) momentum fraction $z$, we need to study the distributions in terms of experimentally-accessible quantities. As discussed in Refs. \cite{deFlorian:2010vy,Renteria-Estrada:2022scipost}, it is possible to obtain approximations to the momentum fractions, based on a proper reconstruction and estimation of the parton-level kinematics. Due to the fact that this analysis is focused on theoretical simulations, we have access to the $z$ variable generated by the MC integrator ($z_{REAL}$), and we can compare the results with possible reconstructions using the kinematics of the external particles ($z_{REC}$). In particular, we used 
\begin{equation}
    \label{eq:zrec}
    z_{REC}=\frac{p_T^h}{p_T^\gamma}\,,
\end{equation}
because it leads to a simpler implementation and the results obtained with the other approximations described in Refs. \cite{deFlorian:2010vy,Renteria-Estrada:2022scipost} are rather similar. Our analysis is restricted to the region $z\in(0.1,\,0.8)$, due to two main reasons. On one side, the cross-section is strongly suppressed outside this range because of kinematical constraints. On the other, the FFs are reliable only in that range, so any attempt to extrapolate them might lead to unphysical conclusions.
\begin{figure*}[t!]
    \centering
    \includegraphics[width=0.49\textwidth]{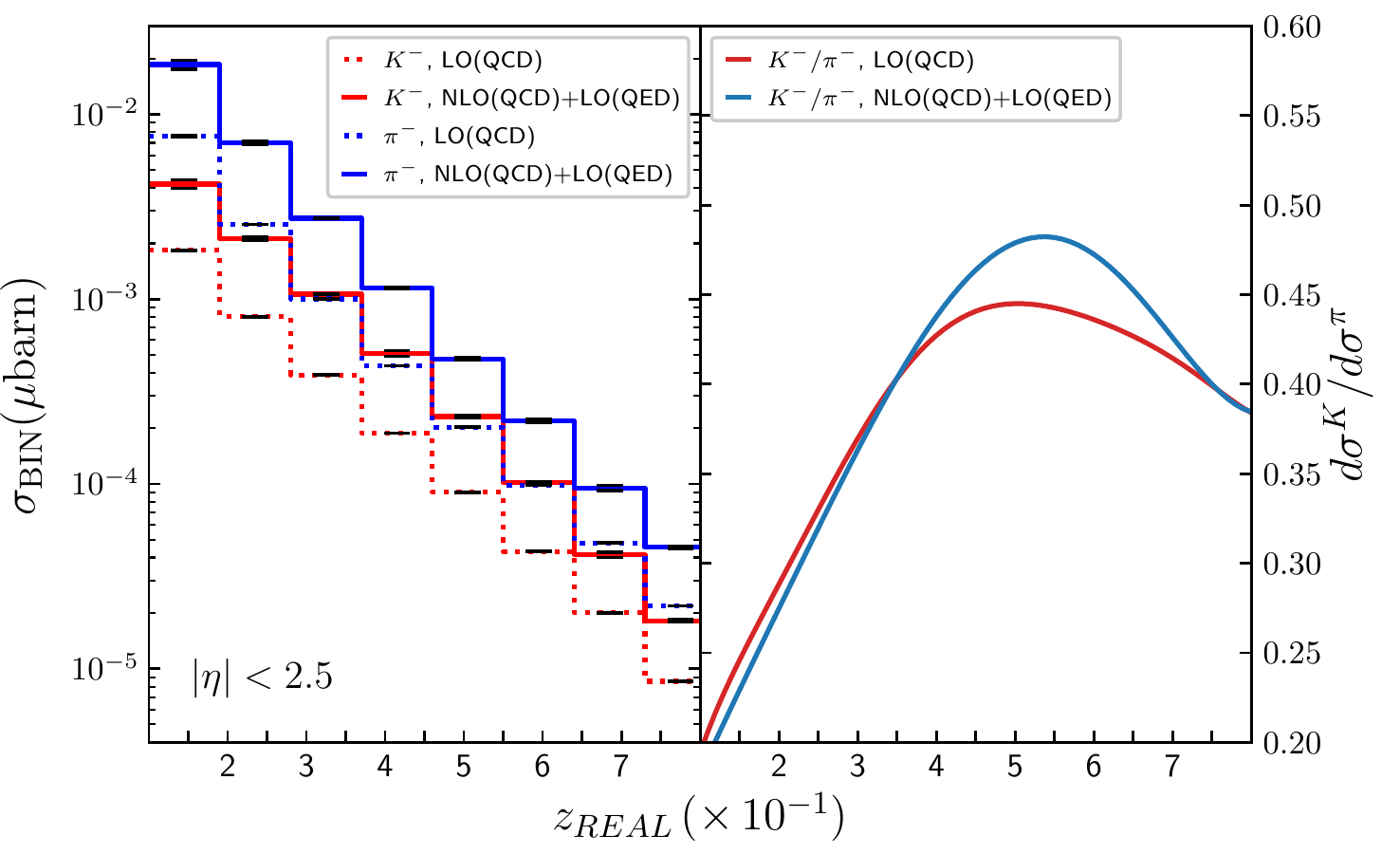}
    \includegraphics[width=0.49\textwidth]{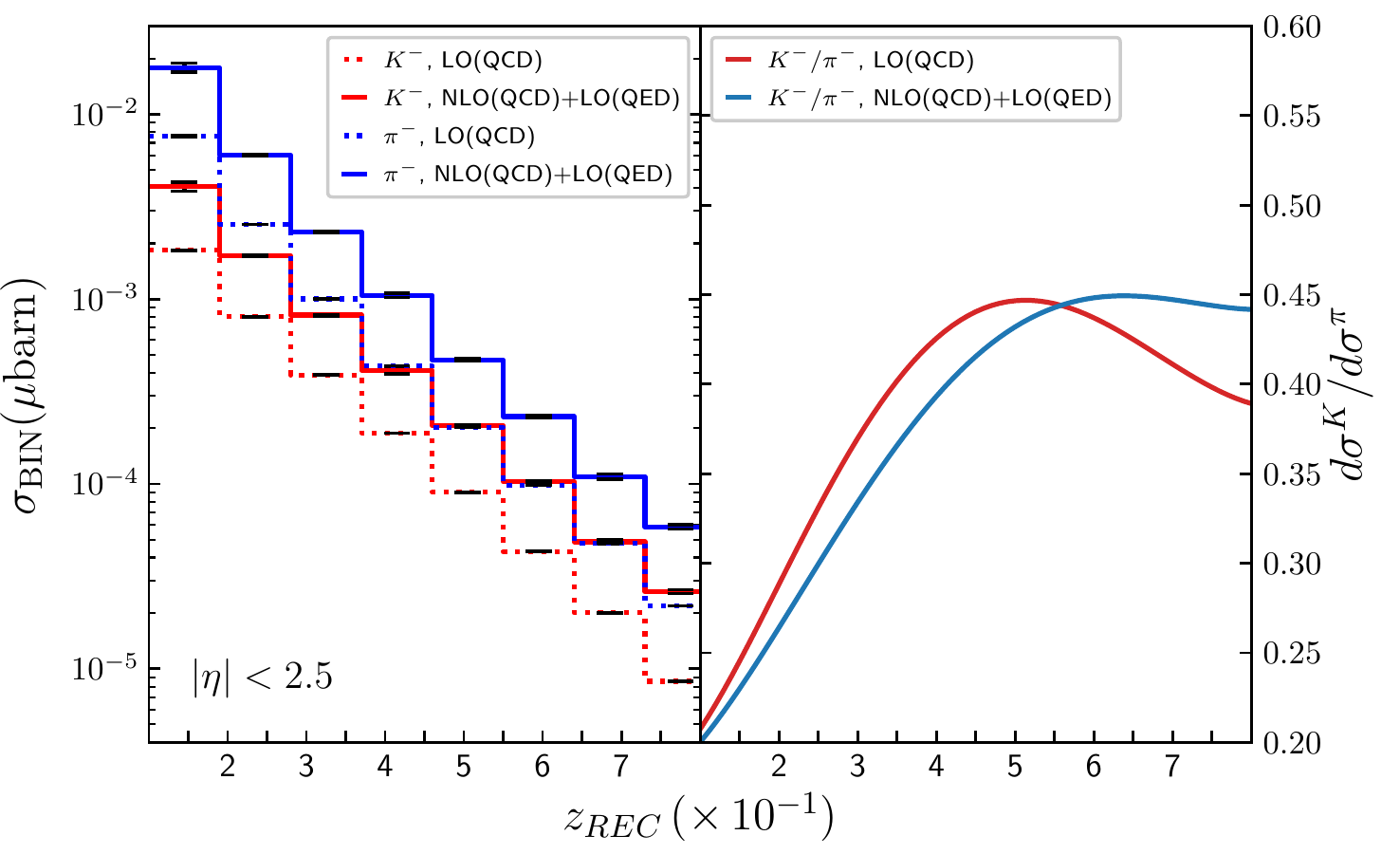}
    \includegraphics[width=0.49\textwidth]{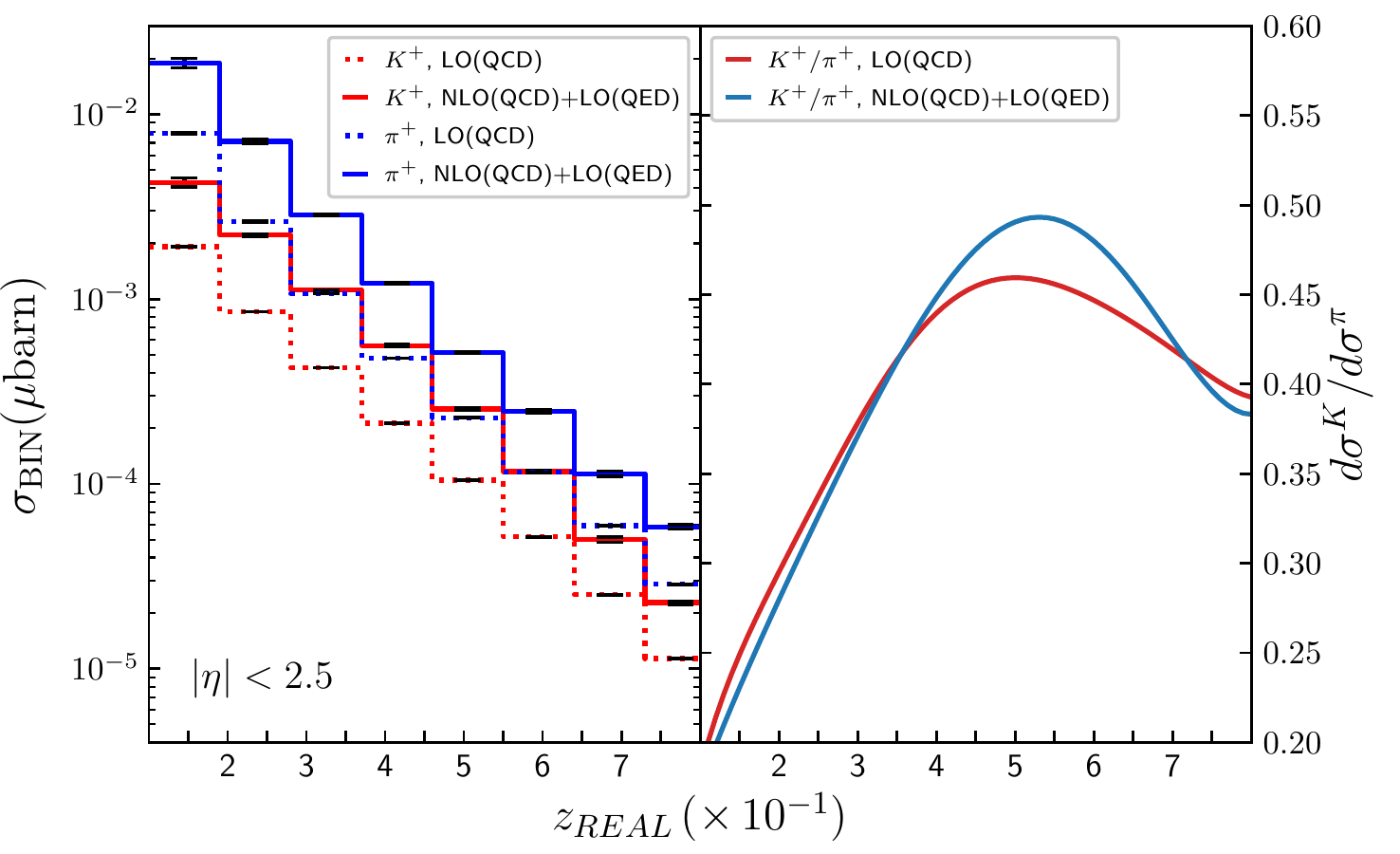}
    \includegraphics[width=0.49\textwidth]{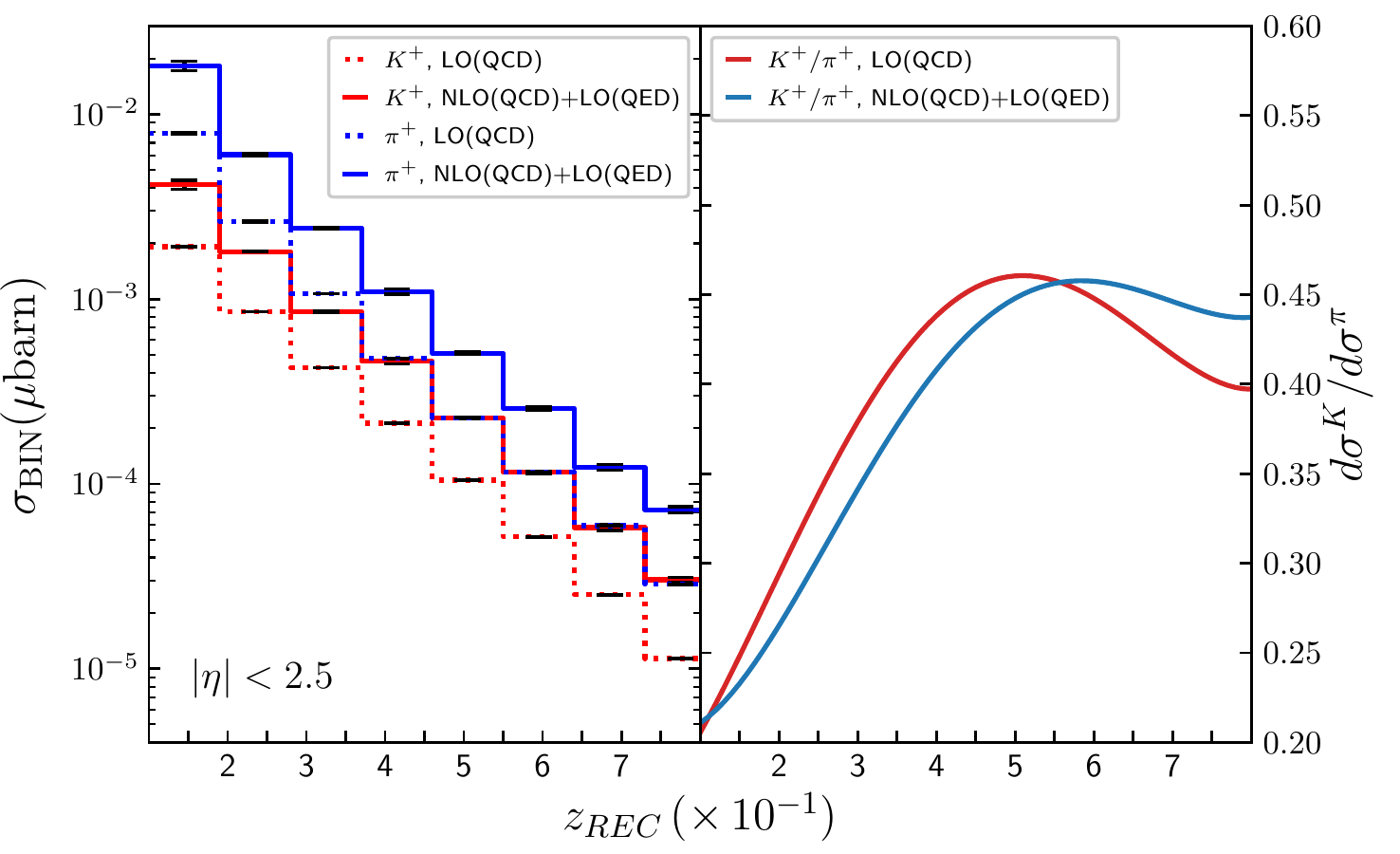}
    \includegraphics[width=0.49\textwidth]{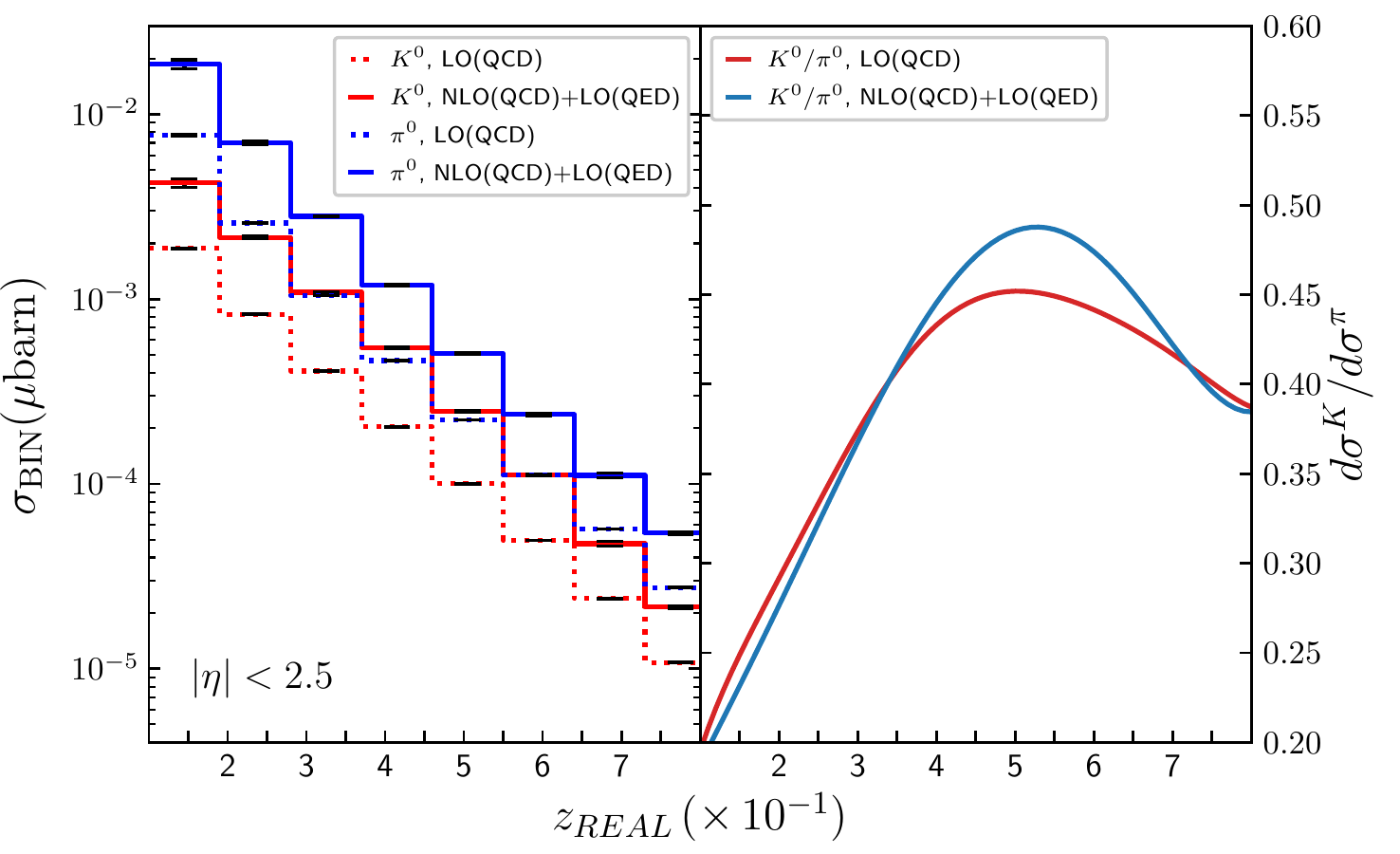}
    \includegraphics[width=0.49\textwidth]{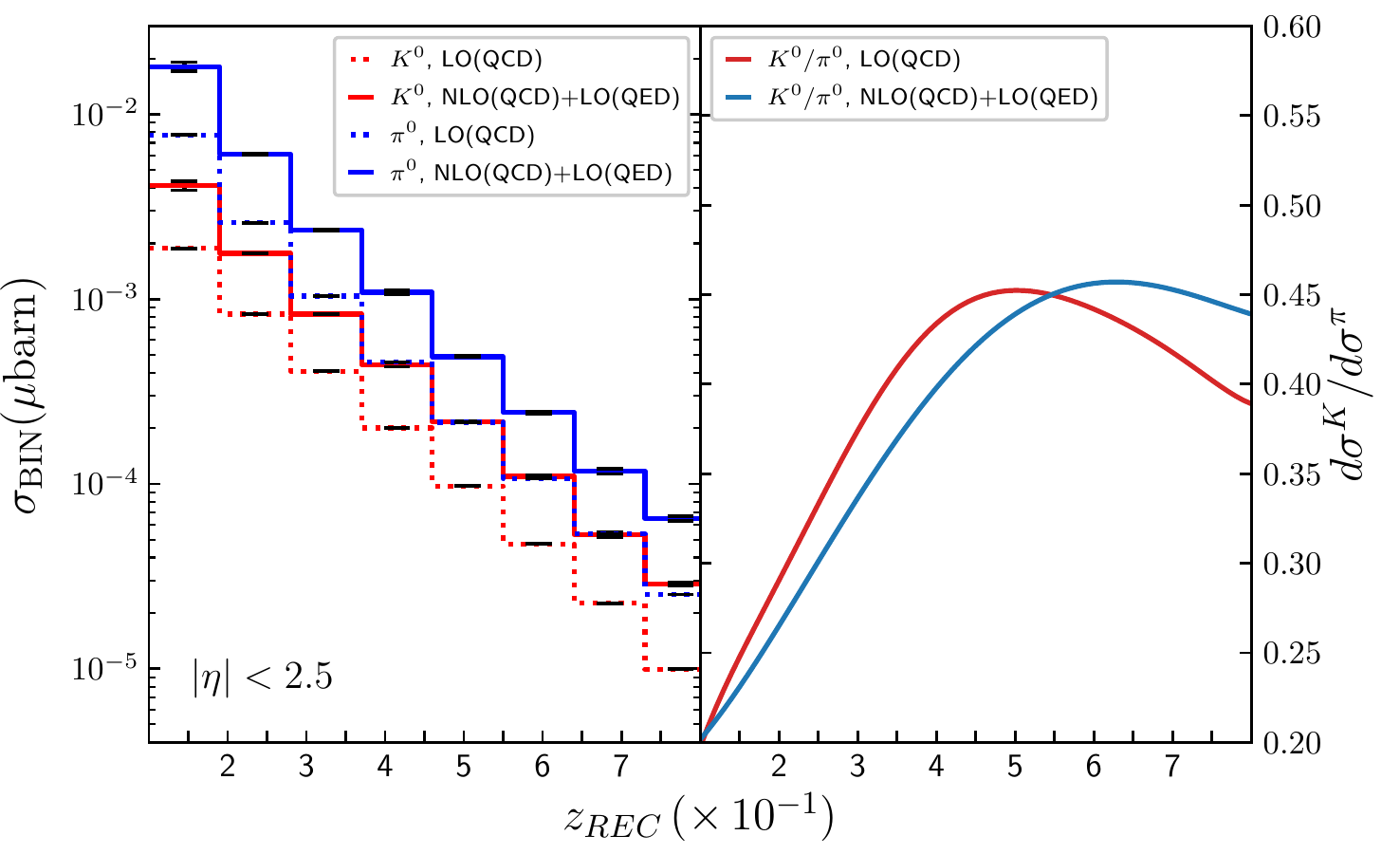}
    \caption{Differential cross-section distribution of $\pi$ and $K$ as a function of $z_{REAL}$ (left column) and $z_{REC}$ (right column), using LHC kinematics ($\sqrt{S_{CM}} = 13 \, {\rm TeV}$). In the left panels, we display the differential cross-sections, whilst in the right ones we present the ratio $d\sigma^K/d\sigma^\pi$. We present the results for negative (first row), positive (second row) and neutral (third row) hadron production. }
    \label{fig:Figura1}
\end{figure*} 

Before moving into the analysis, we would like to motivate the fact that we compared the $z$-spectrum of $\gamma + \pi^{+ \, (-)}$ vs. $\gamma + K^{+ \, (-)}$, but also considered $\gamma + \pi^0$ vs. $\gamma + K^0$. Since the purpose of the study is to relate pion and kaon FFs, we expect $K^{+ \, (-)}$ and $\pi^{+ \, (-)}$ to share similarities because they have the same electromagnetic and isospin charges. However, whilst $\{ \pi^+,\pi^0,\pi^- \}$ belongs to the same isospin multiplet, it is not true for $\{K^+,K^-\}$ and $K^0$. In any case, since $K^0$ and $\pi^0$ are both neutral, we expect to reduce (or cancel) any electromagnetic deviation effect when comparing $\gamma + \pi^0$ and $\gamma + K^0$. As we will better explain later in the article, our methodology, in fact, turns out to be independent of the nature of the hadrons that we are comparing.

In Fig. \ref{fig:Figura1} we display the differential cross-section results w.r.t. $z_{REAL}$ (left column) and $z_{REC}$ (right column) for both $K$ (red) and $\pi$ (blue). In these plots, we include the LO QCD (dashed line) and NLO QCD + LO QED (solid line) corrections, to take into account higher-order effects. As discussed in Refs. \cite{deFlorian:2010vy,Renteria-Estrada:2021rqp}, the K-factor due to NLO QCD corrections can be higher than 2 (i.e. a 100 $\%$ effect w.r.t. the Born level contribution). Besides, it is important to mention that we obtain rather similar results when considering negative (first row), positive (second row) and neutral (third row) hadron production. This is because the FFs involving quarks are very similar, independently of producing same or opposite sign hadrons, and the different production channels are entangled. In fact, the shapes are almost identical at LO; the only noticeable effect due to NLO corrections is that the peak becomes slightly narrower for negative hadron production.

As expected, we appreciate that $\sigma^K$ is smaller than $\sigma^\pi$ in the left panel of Fig. \ref{fig:Figura1}. This is because kaons are heavier than pions, being their production less favorable: this effect is incorporated \emph{exclusively} within the corresponding FFs. When studying the cross-section rations, in the right panels, there is a compensation of NLO effects, because their contribution is embodied within the partonic cross-section (which partially cancels in the ratio). Still, we notice that $d\sigma^K/d \sigma^\pi$ goes from 0.2 for $z_{REAL}=0.2$ to approximately 0.40 for $z_{REAL}=0.8$, reaching a peak of $0.5$ for $z_{REAL} \approx 0.5$. This behaviour is mainly due to the FFs shapes, but the presence of NLO QCD corrections enhance the peak. In all the cases, the effect of QED corrections is negligible.

Some comments about the difference among the left and right columns of Fig. \ref{fig:Figura1} are worth. In first place, we recall that $z_{REAL}$ is not accessible in the experiments, but we can estimate the shape from our MC simulations. By construction, $z_{REC}$ tends to mimic the behaviour of $z_{REAL}$, specially close to Born-level kinematics. So, we expect that the agreement between both spectra tend to increase as $|\eta|\to 0$. In any case, the distributions in $z_{REC}$ lead to smoother ratios, which are better behaved in the high-$z$ region. Also, the peak around $z=0.5$ is softened specially when dealing with the NLO QCD corrections. This behaviour is expected because we are using a reconstruction formula which is only exact at LO, and the region $z \approx 0.5$ has an important contribution of events including real-radiation corrections (i.e. NLO effects). In fact, this is the reason for the ${\cal O}(10 \%)$ deviation between NLO and LO predictions in the left column (i.e. for the $z_{REAL}$ spectrum).


\subsection{Analysis of different parton-initiated contributions}
\label{ssec:Subchannels}
The second study that we performed is intended to achieve a better understanding of the different partonic contributions involved in the hadron-photon production process. Besides identifying the dominant one, we want to perform a rough estimation of its dependence w.r.t. some kinematical cuts.

\begin{figure}[h!]
    \centering   
    \includegraphics[width=0.49\textwidth]{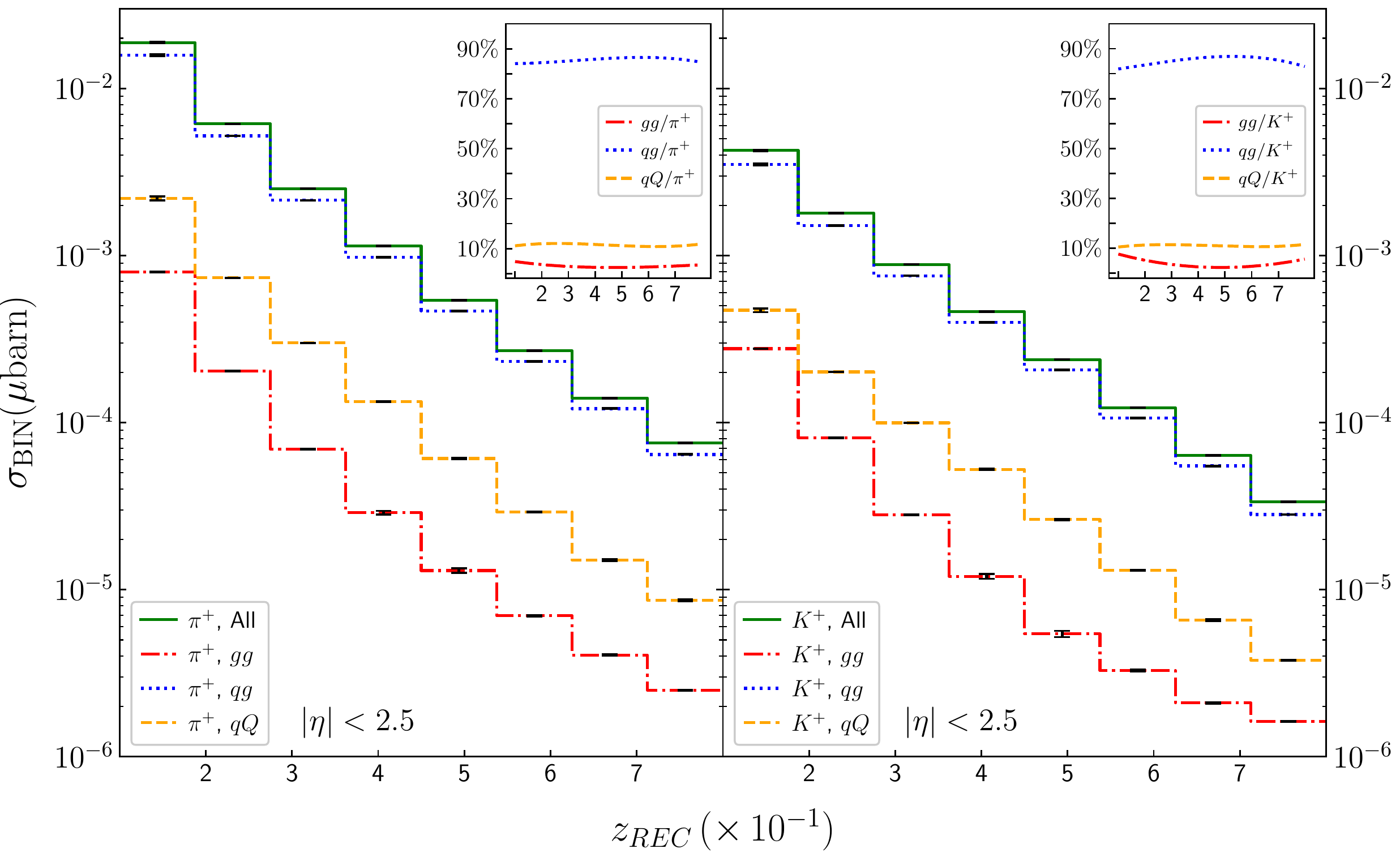}
    \includegraphics[width=0.49\textwidth]{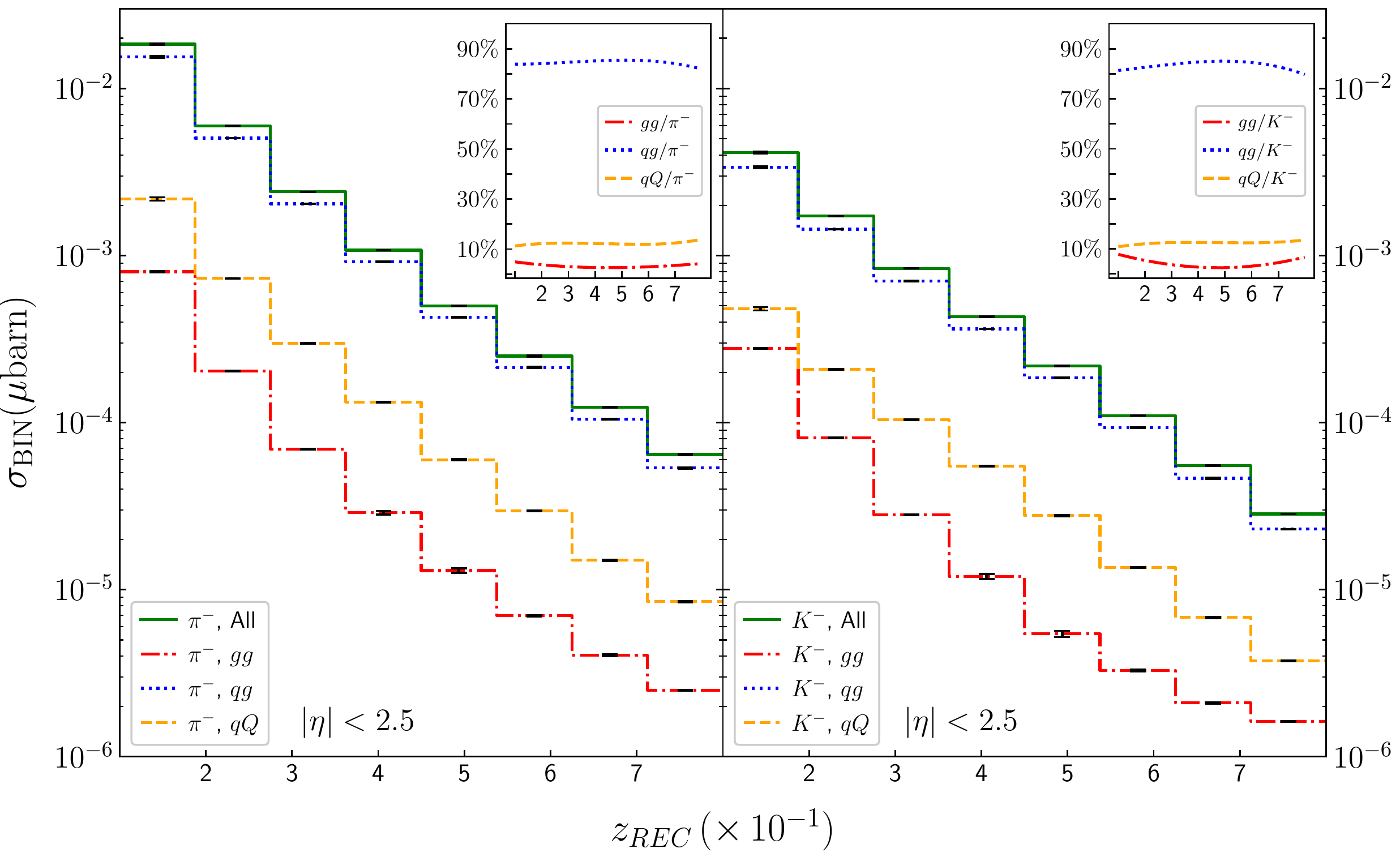}
    \caption{Differential cross-section distribution as a function of $z_{REC}$ for $\pi$ (left) and $K$ (right) production. We indicate the contributions due to the different partonic channels: $gg$ (dashed red line), $qg$ (point blue line) and $qQ$ (dashed orange line). We use LHC kinematics ($\sqrt{S_{CM}} = 13
    \, {\rm TeV}$) and present the results for positive (upper row) and negative (lower row) hadron production. In the upper sector of each plot, we indicate the relative contribution of each channel.}
    \label{fig:Figura2}
\end{figure}

In Fig. \ref{fig:Figura2}, we present the contributions of the different partonic channels to the cross-section w.r.t. $z_{REC}$, for pion (left panel) and kaon (right panel) production. We use the LHC cuts specified in the beginning of this Section, and we include predictions for positive (upper row) and negative (lower row) hadron production, up to NLO QCD + LO QED accuracy. From these plots, we reconfirm that pion is enhanced w.r.t. kaon production, but the contributions of the different partonic channels is almost the same. Also, we find that the cross-section rates are roughly identical independently of the charge of the produced hadron. The most important conclusion from these plots is that $qg$ channel dominates, being a factor ${\cal O}(10)$ larger than the others. It is also interesting to notice that $qQ$ channel is larger than $gg$ one, even including the gluon PDF enhancement for proton-proton collisions. However, this difference tends to reduce for larger values of $z$. Regarding $q\gamma$ channel, it is greatly suppressed compared to the NLO QCD contributions; for this reason, we are not displaying it in Fig. \ref{fig:Figura2}.

\begin{figure}[h!]
    \centering
    \includegraphics[width=0.49\textwidth]{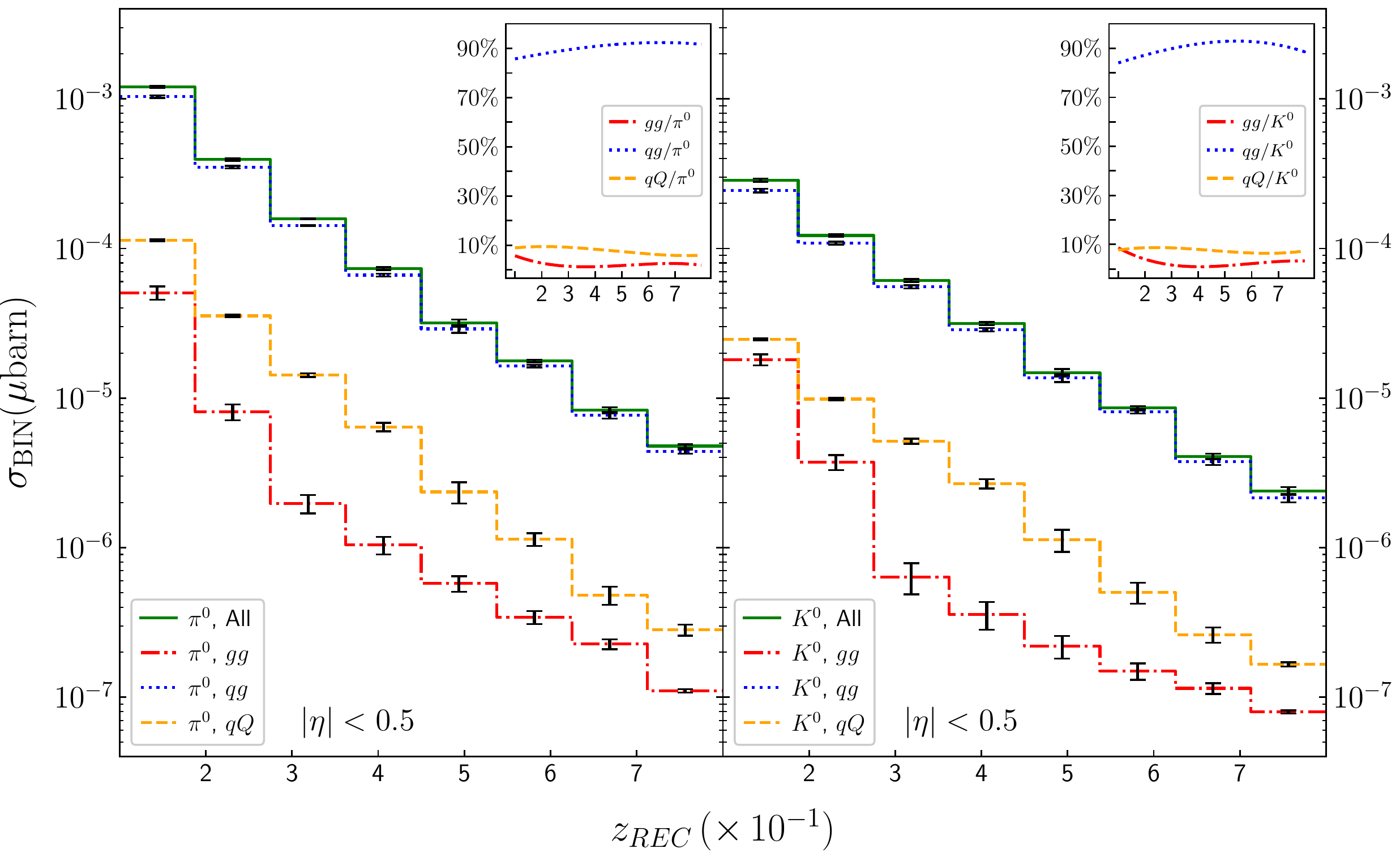}
    \includegraphics[width=0.49\textwidth]{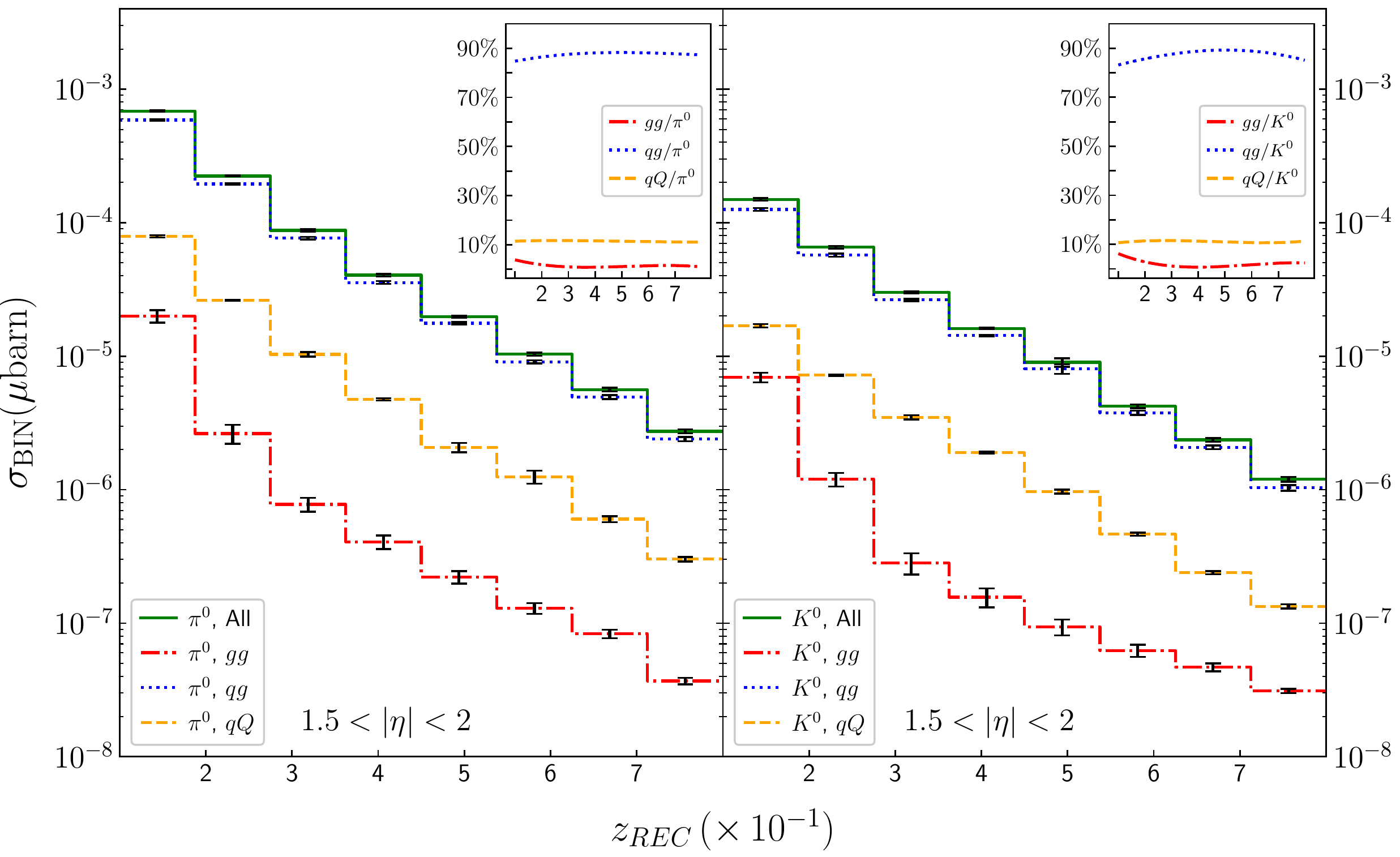}
    \caption{Differential cross-section distribution as a function of $z_{REC}$ with LHC kinematics ($\sqrt{S_{CM}} = 13 \, {\rm TeV}$), for Cases {\it 1} (upper row) and {\it 2} (lower row). In the right panel, we consider neutral kaon production, whilst the left one shows the distribution for neutral pion production. In the upper sector of each plot, we indicate the relative contribution of each channel.}
    \label{fig:Figura3}
\end{figure}

As a first exploratory step, in this article we study the weight of the different channels when changing the pseudorapidity cuts. This is relevant because such selection cuts can be easily modified by experimental analysis. So, we considered the default scenario (Case {\it 0}), described in the beginning of this Section, along with:
\begin{itemize}
    \item{Case {\it 1}: }\quad $\{|\eta^h|,|\eta^\gamma|\}< 0.5$,
    \item{Case {\it 2}:}\quad $1.5<\{|\eta^h|,|\eta^\gamma|\}<2$,
\end{itemize}
keeping all the other variables unchanged. The idea is that Case {\it 1} should retain those events that are closer to the Born-level kinematics, thus favouring $qg$ and $qQ$ channels. On the contrary, Case {\it 2} is expected to enhance the contributions originated by non-Born kinematics. In Fig. \ref{fig:Figura3}, we study the weight of the different contributions for Cases {\it 1} (upper row) and {\it 2} (lower row), and considering neutral pion (left) and kaon (right) production. We decided to focus on the production of neutral hadrons because the distributions are more stable (which can be expected since they are the average of positive and negative distributions). In both Cases, we observe that the dominant channel continues to be $qg$, although there are non-negligible changes in the $qQ$ and $gg$ channels. In the $|\eta|<0.5$ region, the $qQ$ and $gg$ channels are very similar; but the difference is augmented when considering $1.5<|\eta|<2$. This behaviour is rather similar for both pion and kaon production, although the latest shows $qQ$ and $gg$ much closer than for pion production. In any case, this points towards an effect originated at the partonic cross-section level, rather than induced by the FFs.

\begin{figure*}[t]
    \centering
    \includegraphics[width=0.49\textwidth]{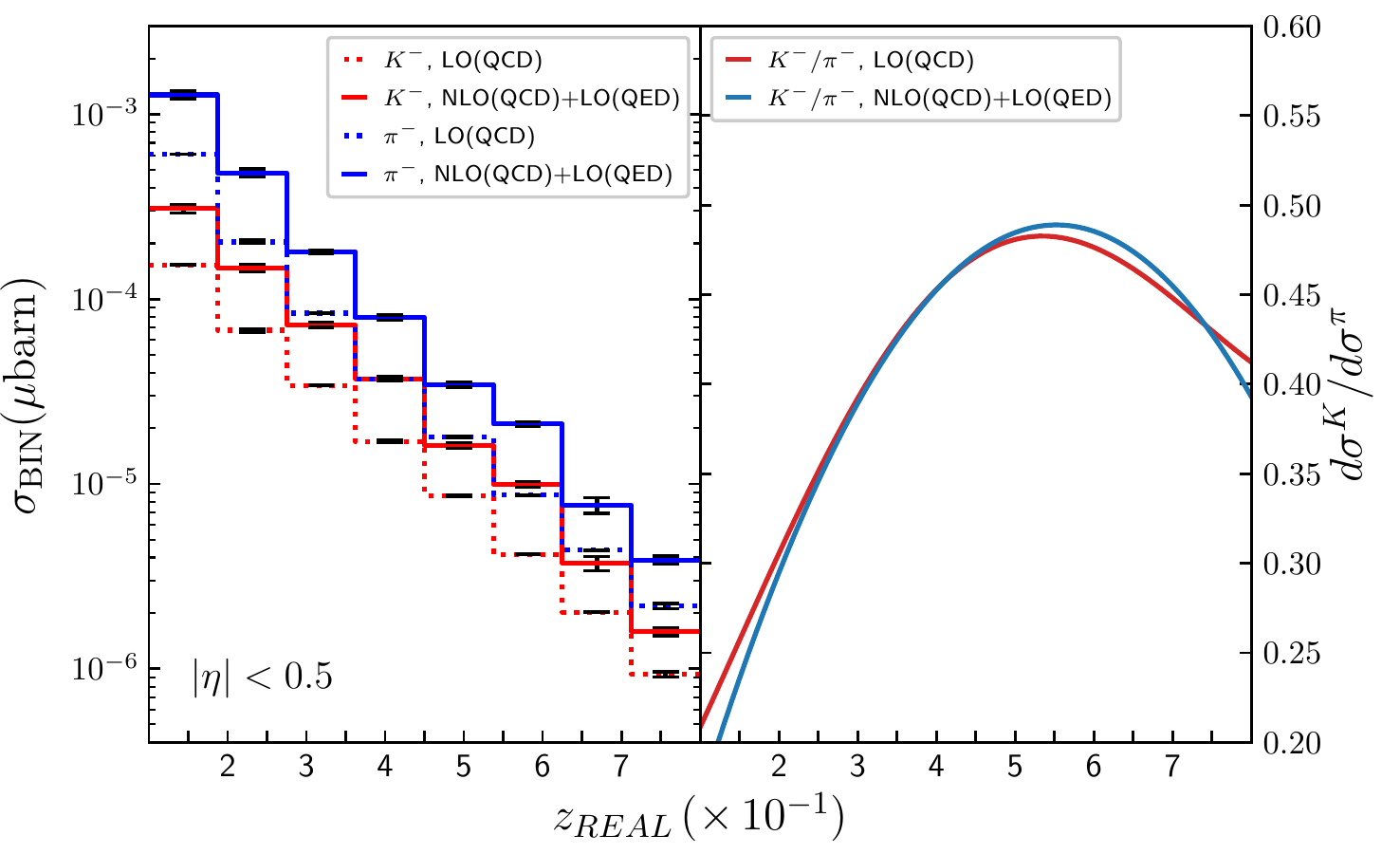}
    \includegraphics[width=0.49\textwidth]{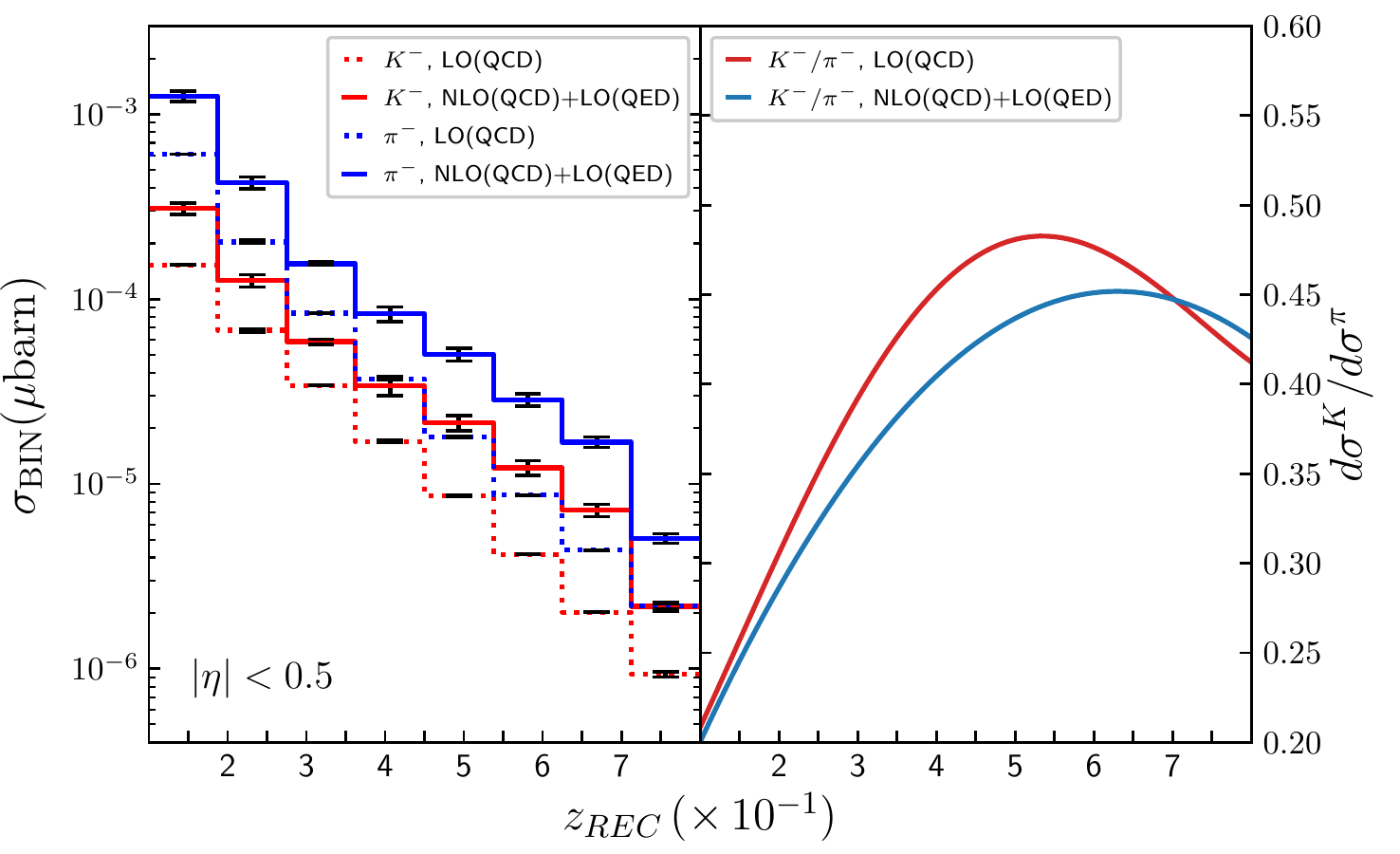}
    \includegraphics[width=0.49\textwidth]{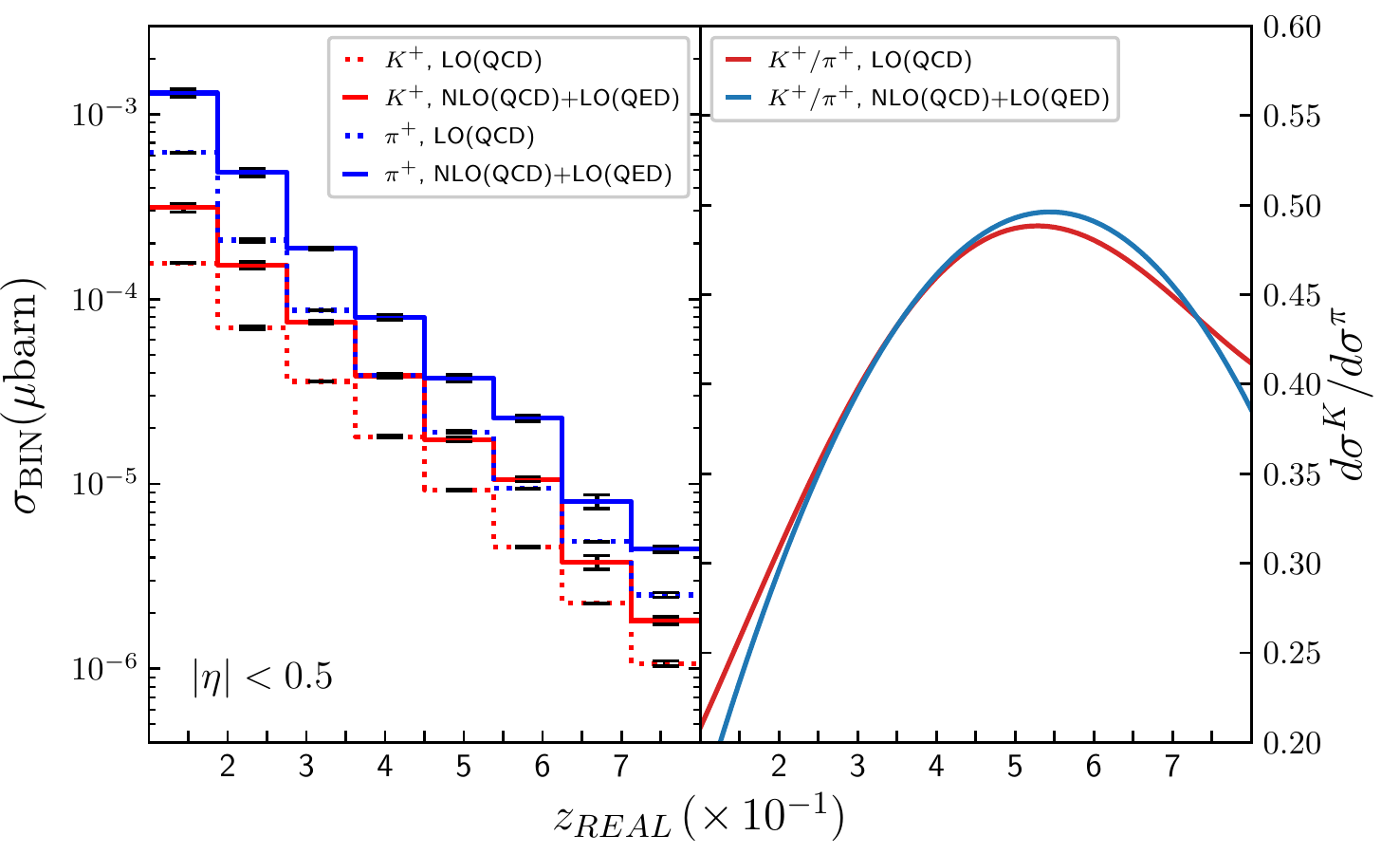}
    \includegraphics[width=0.49\textwidth]{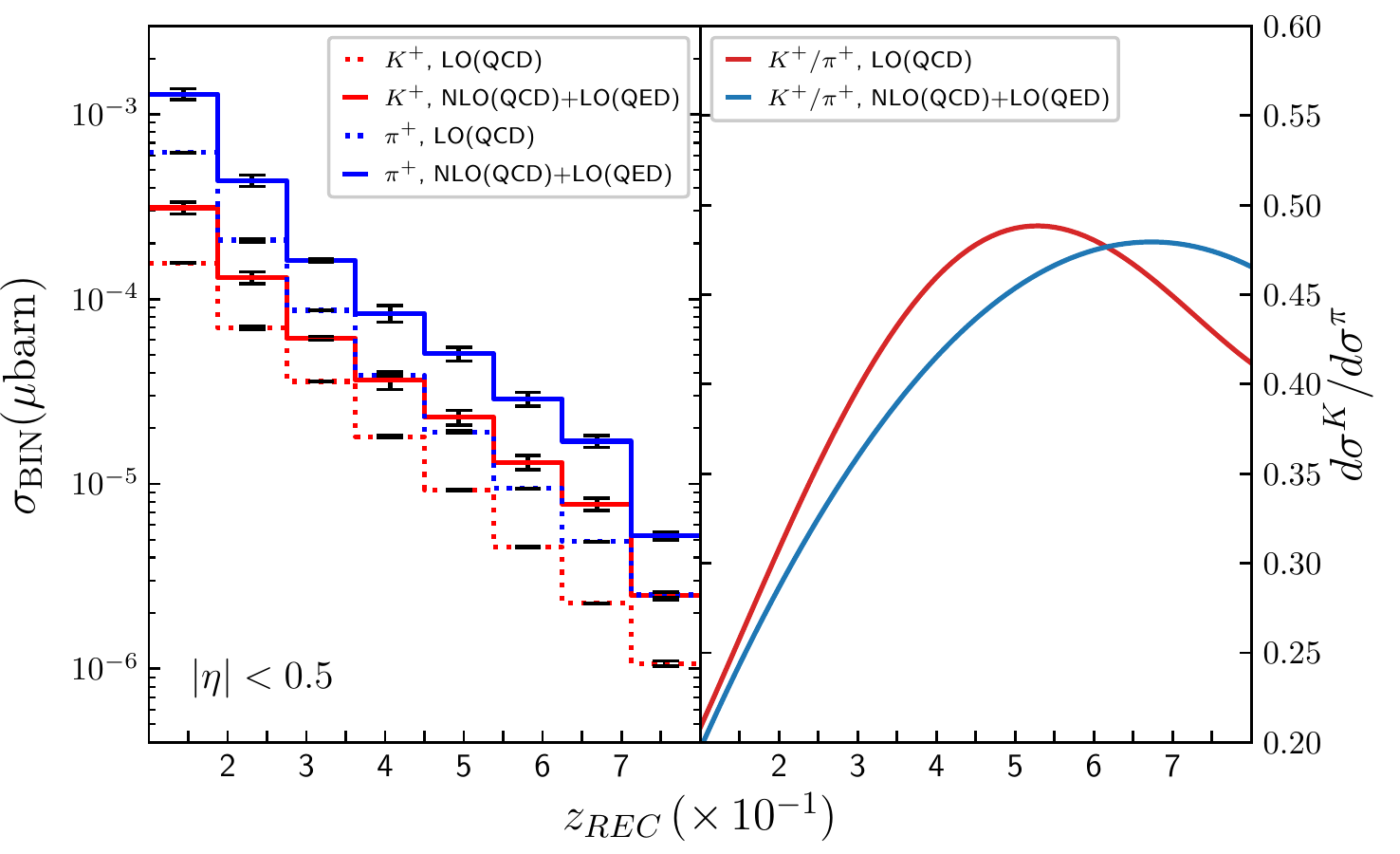}
    \includegraphics[width=0.49\textwidth]{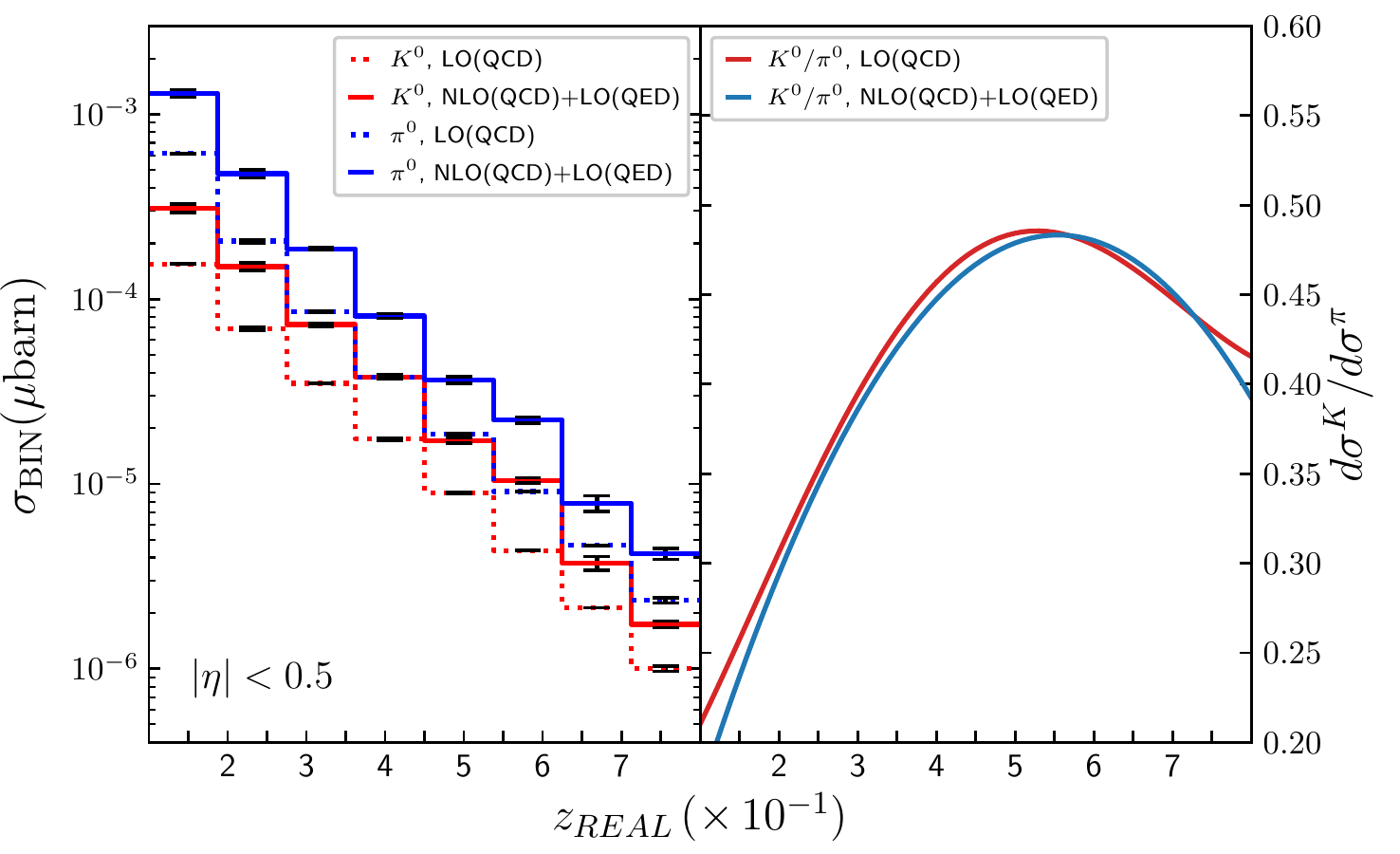}
    \includegraphics[width=0.49\textwidth]{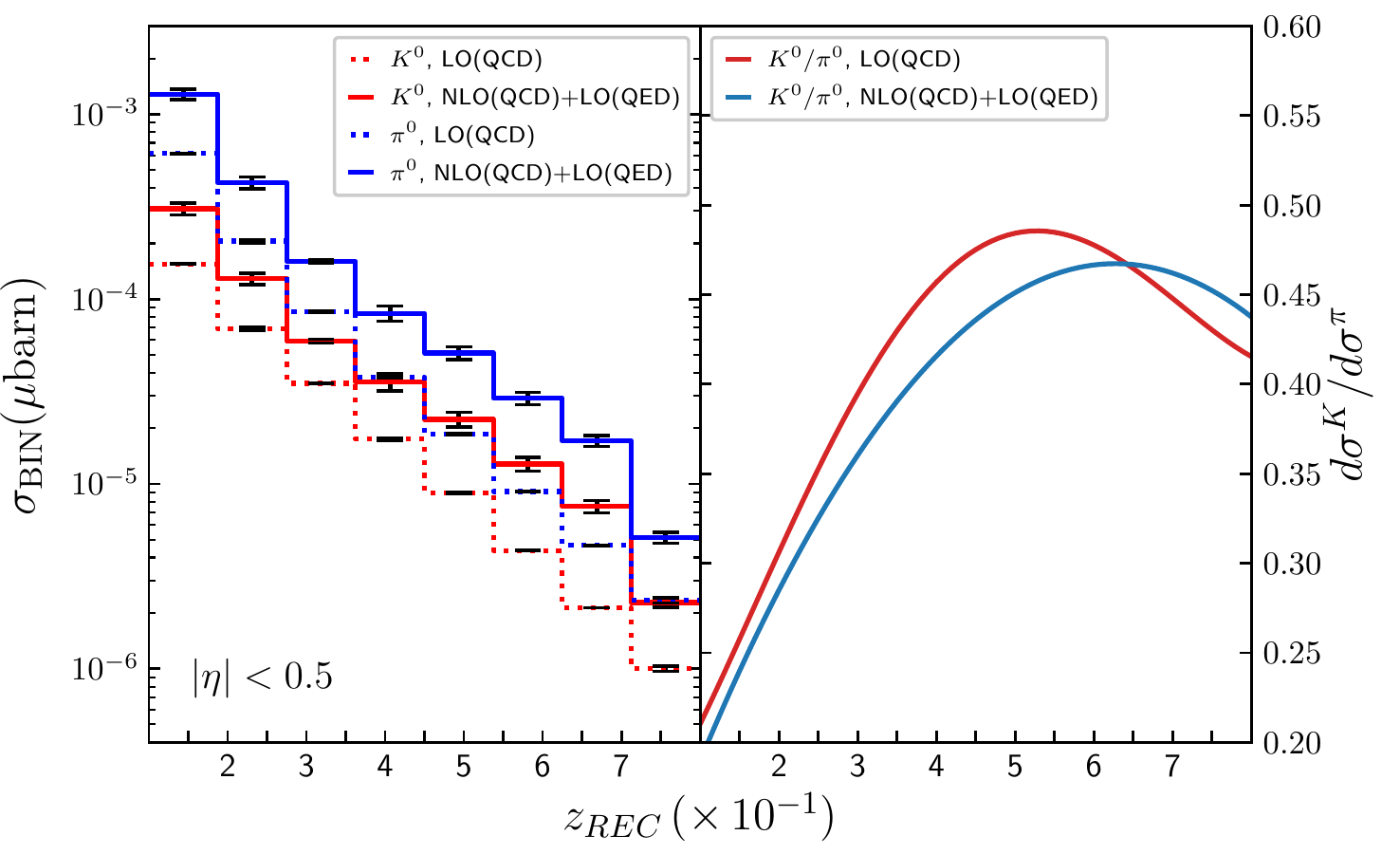}
    \caption{Same as Fig.\ref{fig:Figura1} for hadrons with charge $-1$ (first row), $+1$ (second row) and $0$ (third row), with the kinematical cut $|\eta|<0.5$ (Case {\it 1}).}
    \label{fig:Figura4A}
\end{figure*}

\begin{figure*}[t]
    \centering
    \includegraphics[width=0.49\textwidth]{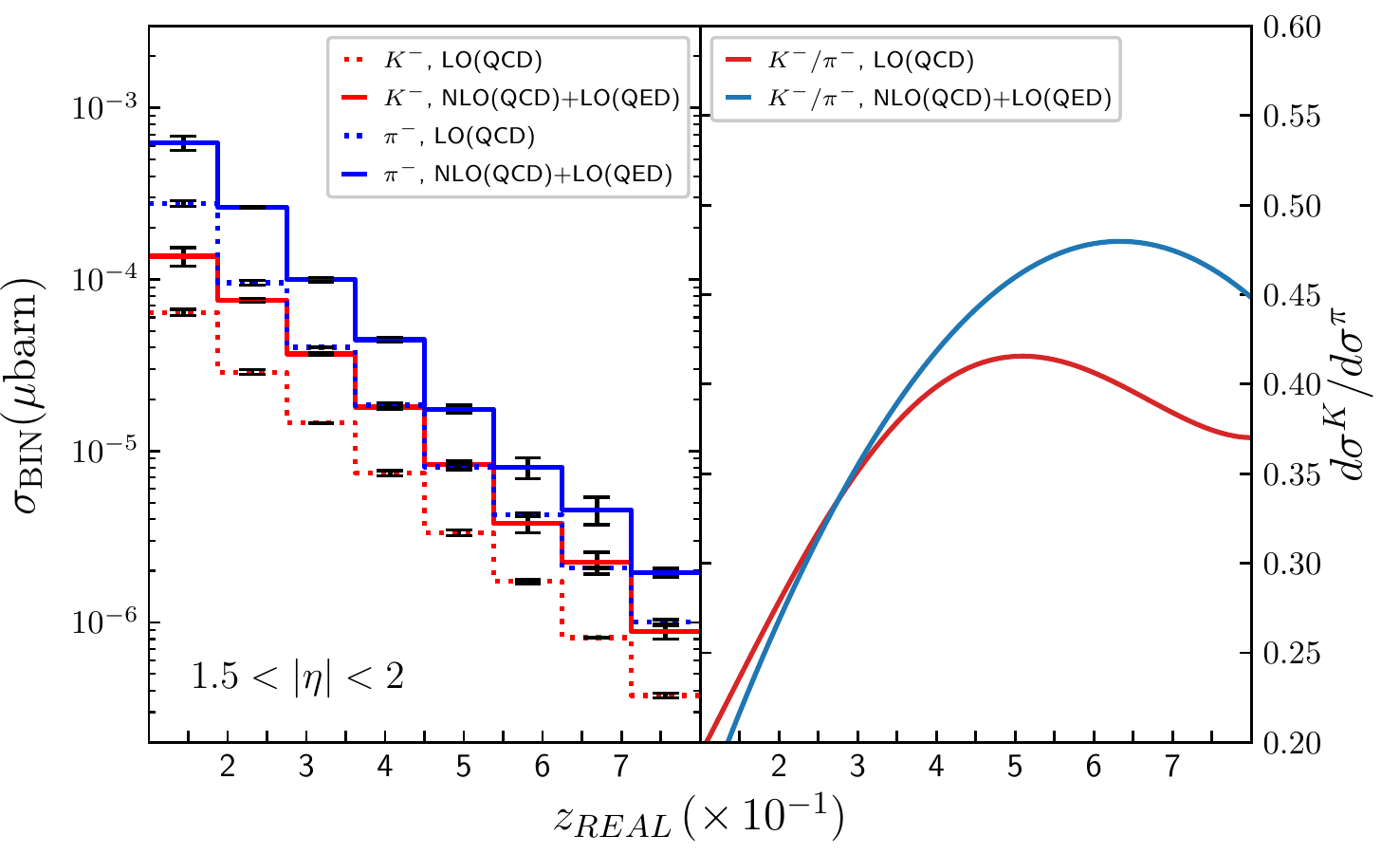}
    \includegraphics[width=0.49\textwidth]{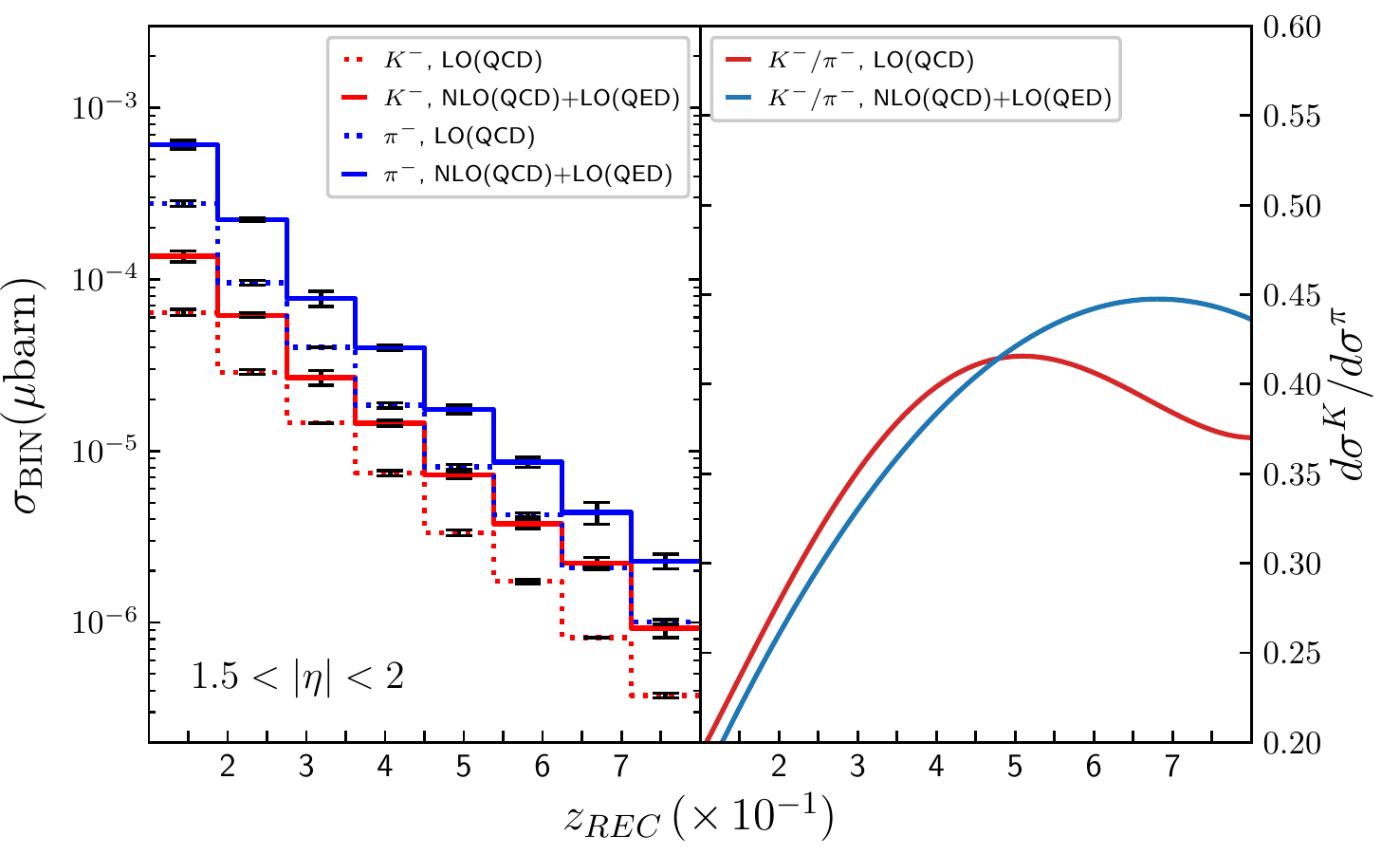}
    \includegraphics[width=0.49\textwidth]{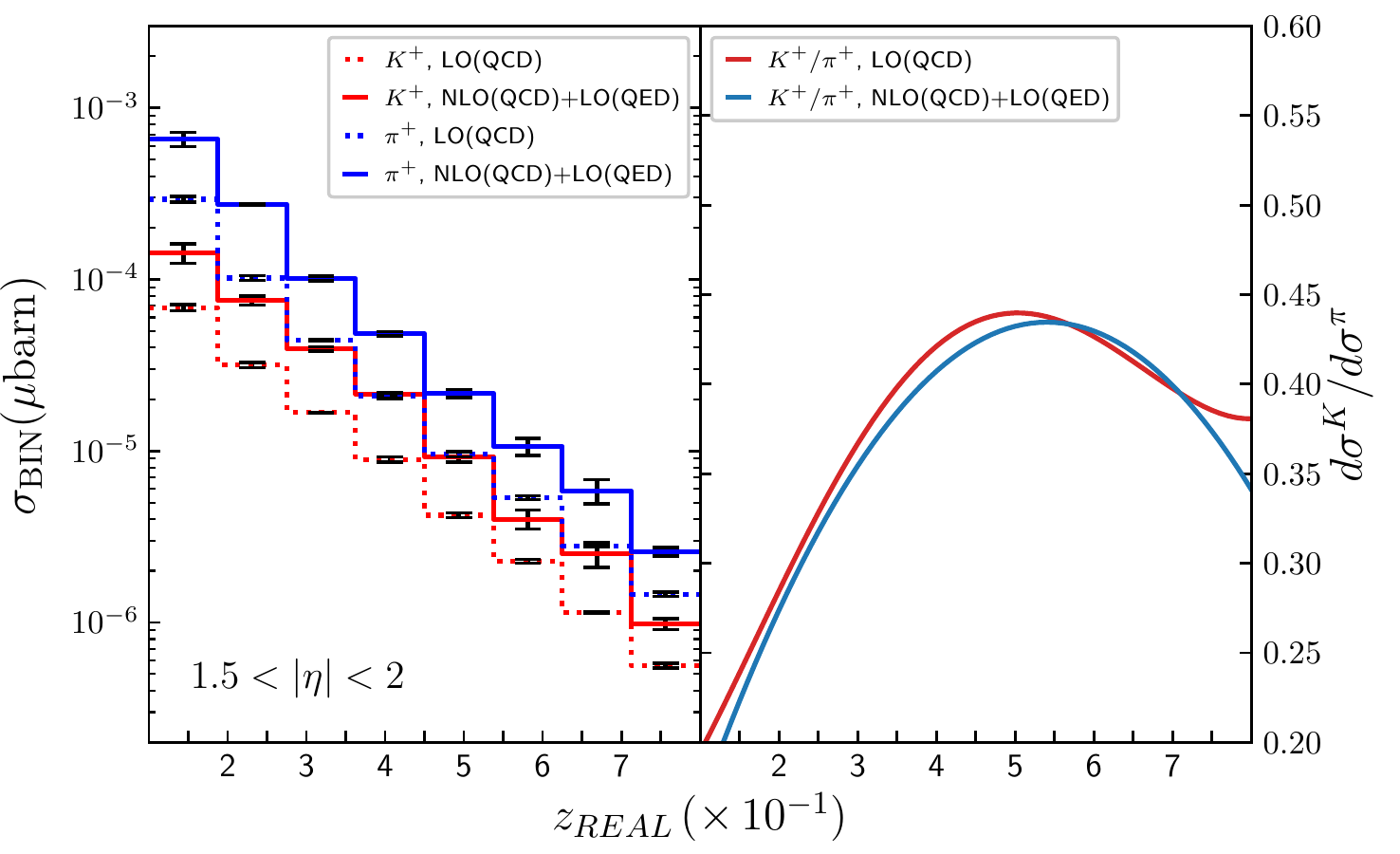}
    \includegraphics[width=0.49\textwidth]{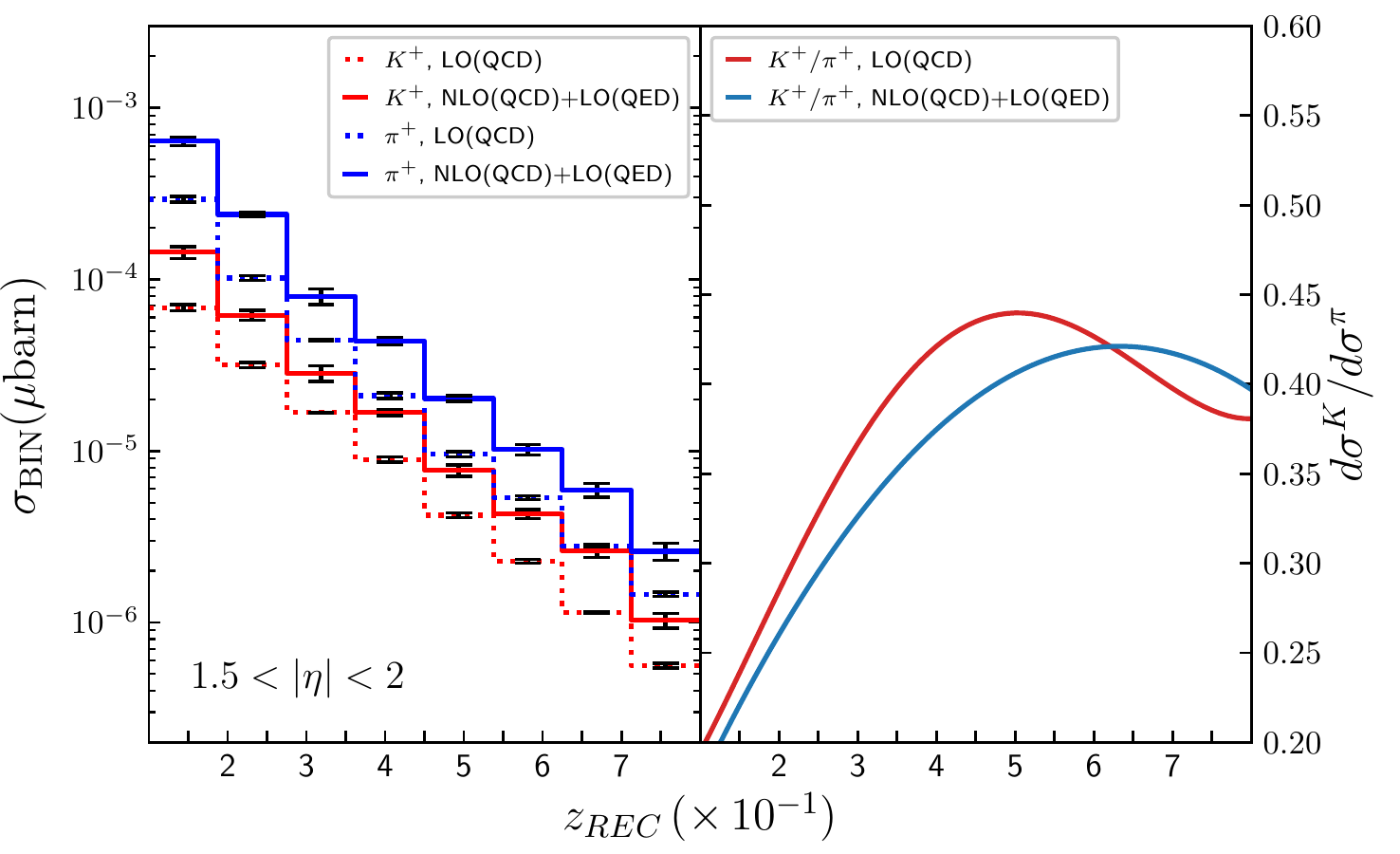}
    \includegraphics[width=0.49\textwidth]{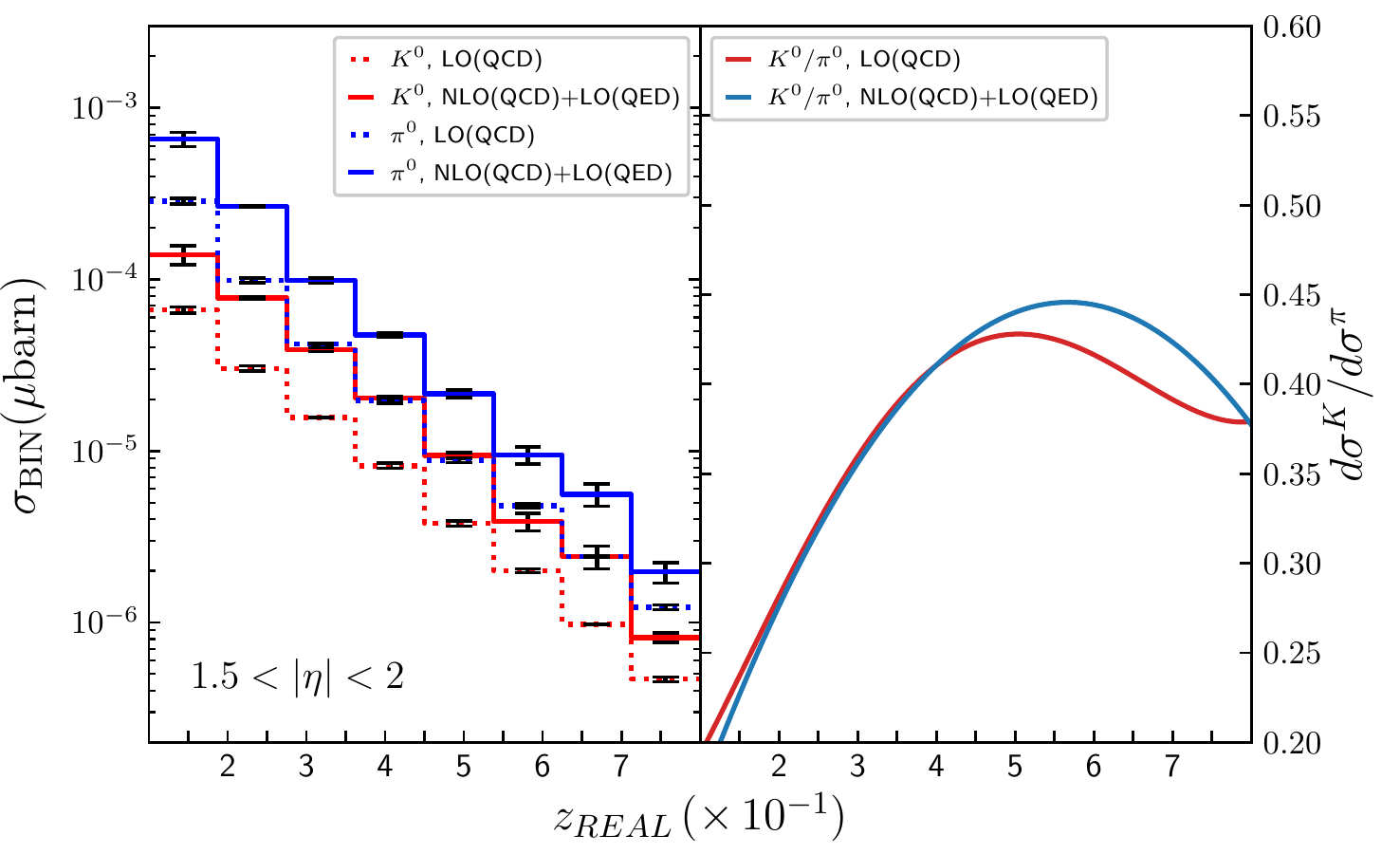}
    \includegraphics[width=0.49\textwidth]{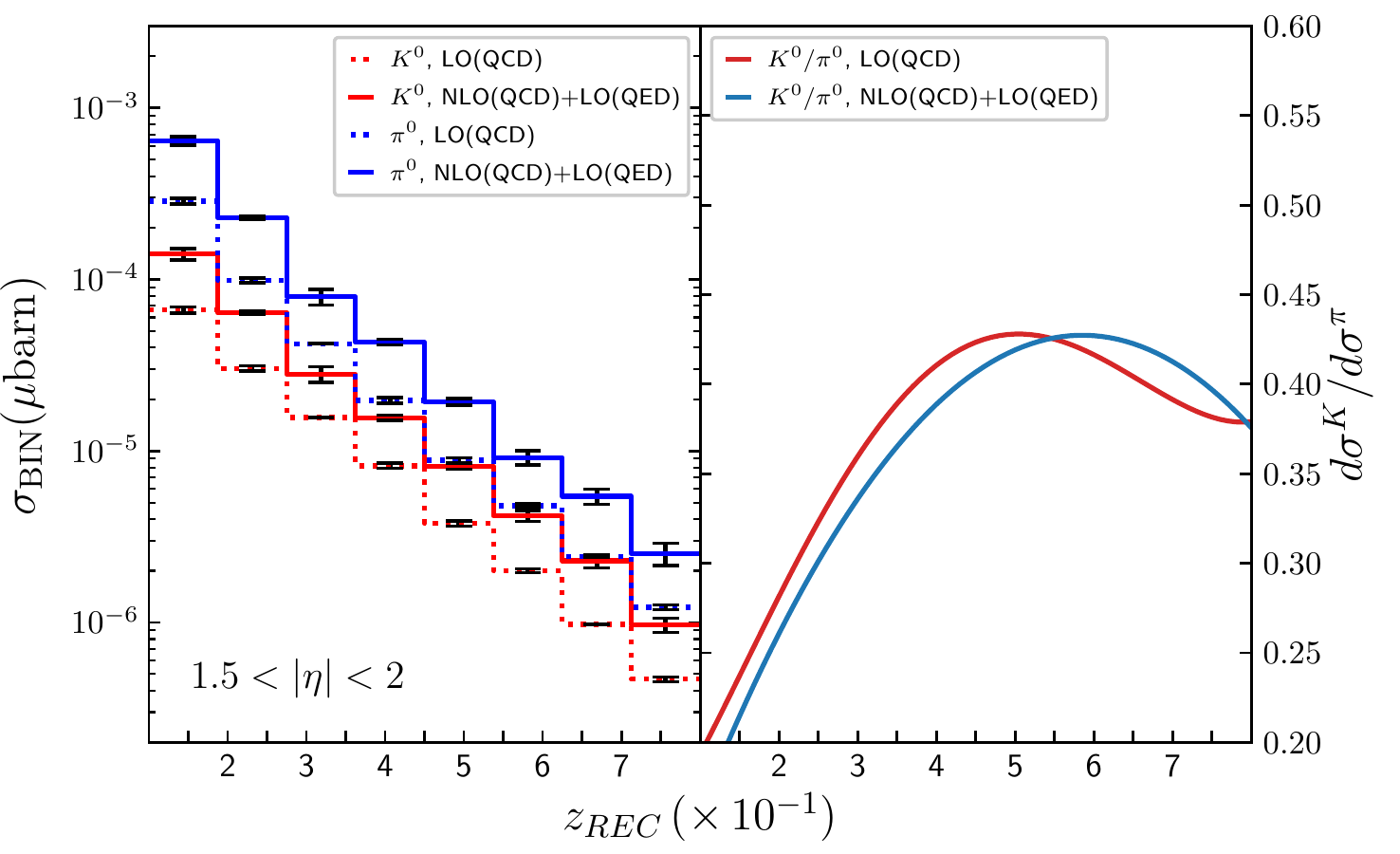}
    \caption{Same as Fig.\ref{fig:Figura1} for hadrons with charge $-1$ (first row), $+1$ (second row) and $0$ (third row), with the kinematical cut $1.5<|\eta|<2$ (Case {\it 2}).}
    \label{fig:Figura4B}
\end{figure*}

\begin{figure}[h!]
    \centering    
    \includegraphics[width=0.49\textwidth]{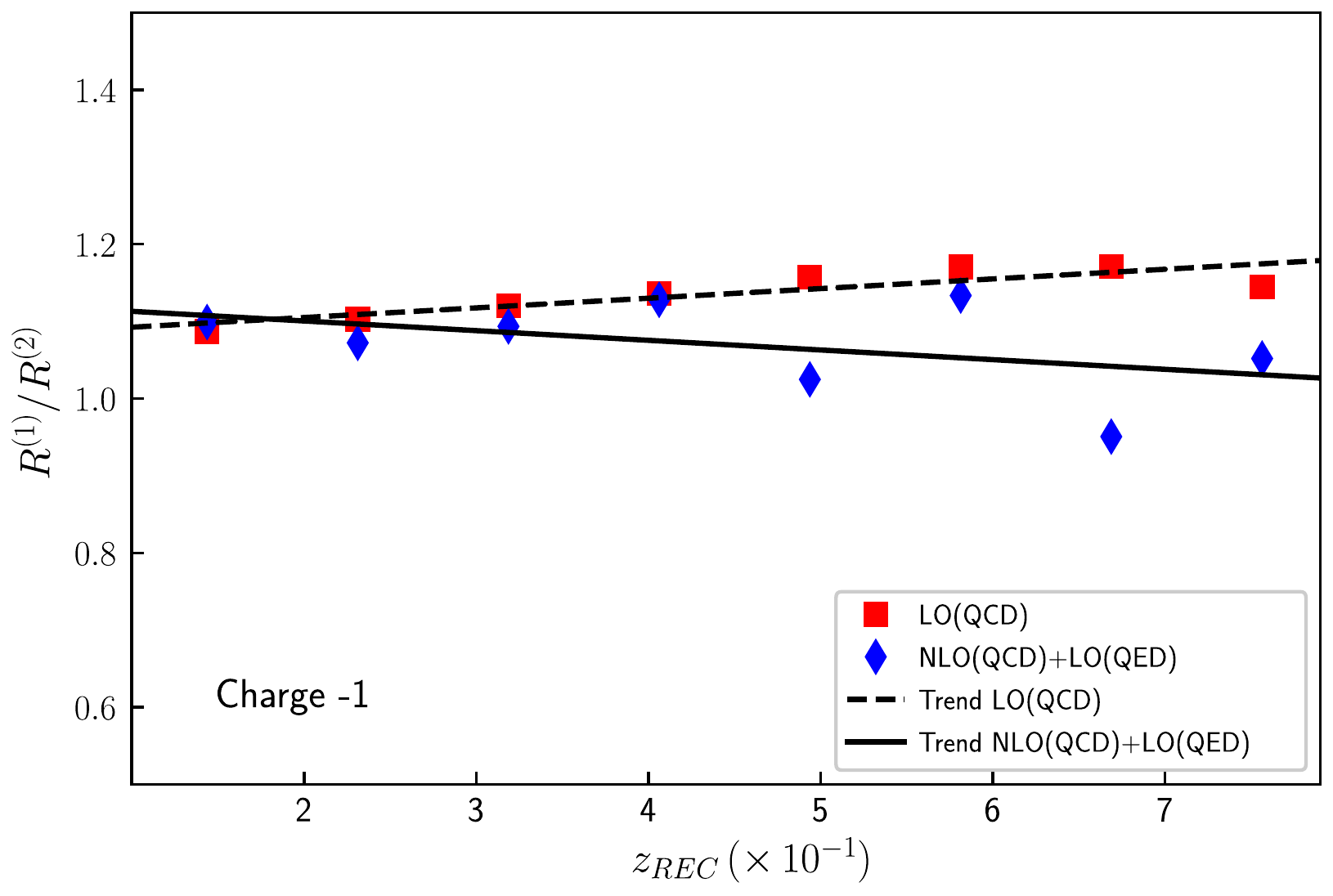}   
    \includegraphics[width=0.49\textwidth]{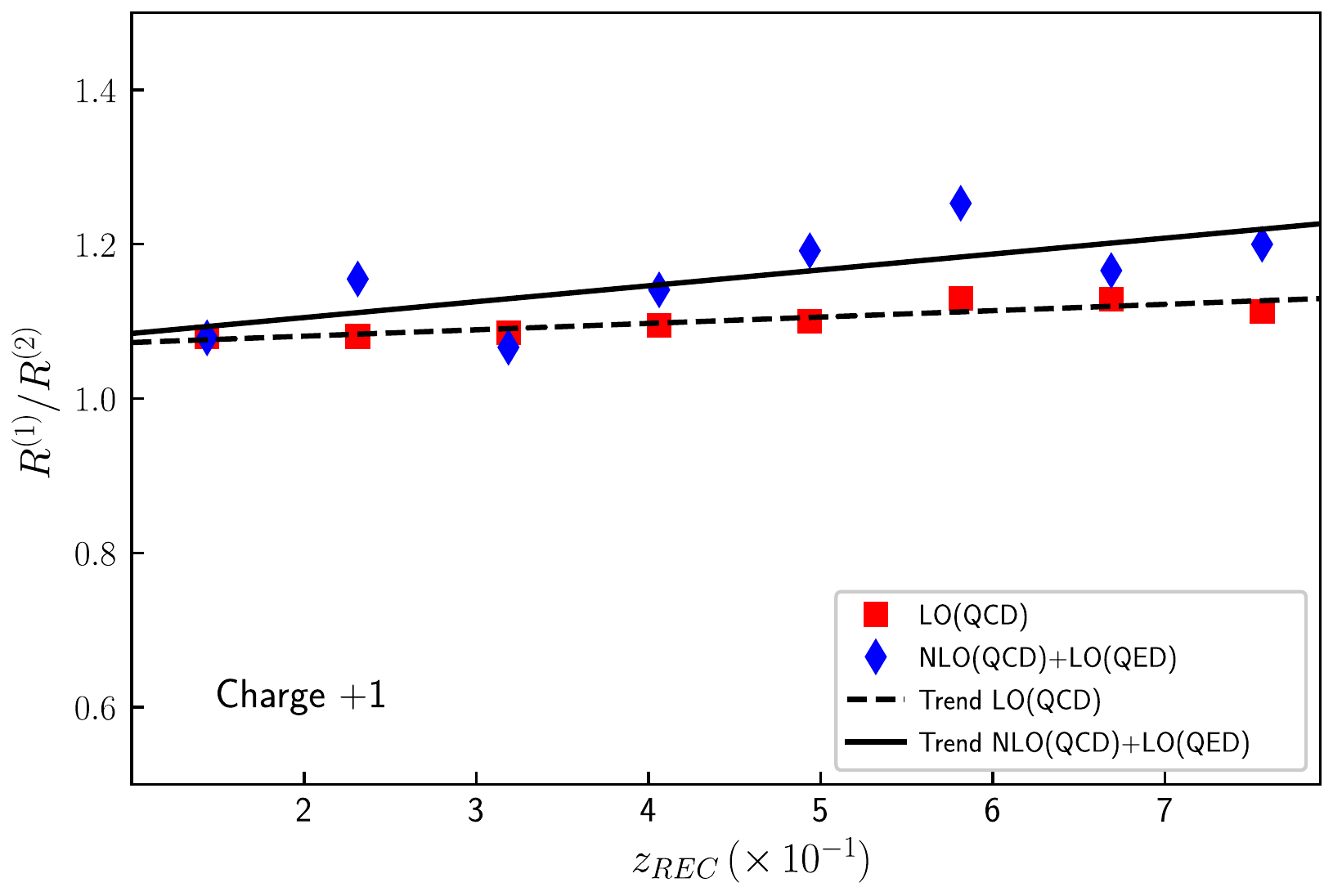}    
    \includegraphics[width=0.49\textwidth]{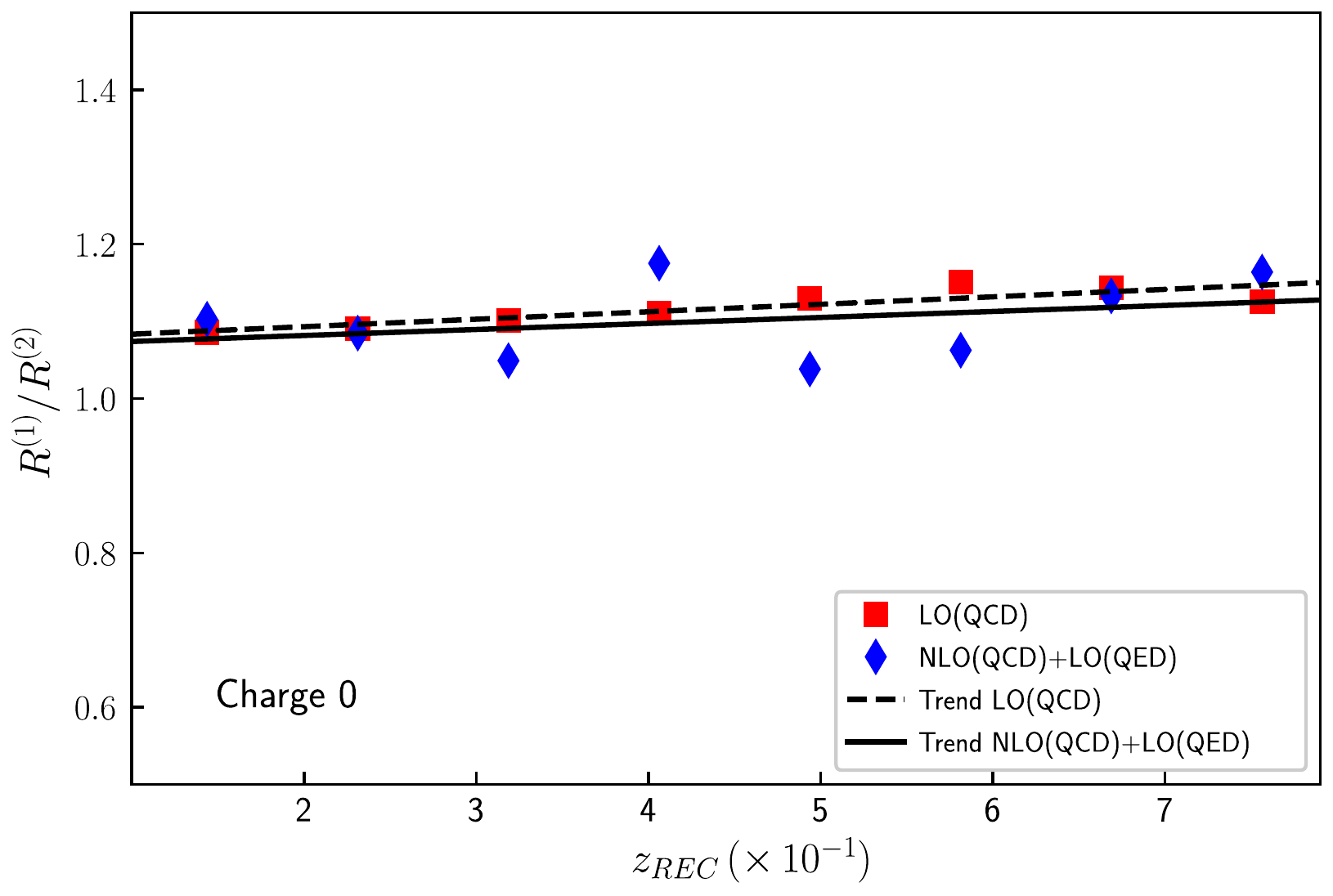}   
    \caption{Comparison of cross-section ratios $R^{(1)}/R^{(2)}$ as a function of $z_{REC}$. We include the LO QCD (red) and NLO QCD + LO QED (blue) predictions, for negative (upper), positive (center) and neutral (lower plot) hadron production.}
    \label{fig:Figura4C}
\end{figure}

Then, we studied the impact of the cuts associated to Cases {\it 1} and {\it 2} in the ratio $d\sigma^K/d \sigma^\pi$, since differences might appear. However, as in the previous case, all the distributions and ratios are rather similar, independently of the charge of the hadron. In Fig. \ref{fig:Figura4A}, we display the differential cross-section distribution as a function of $z_{REAL}$ (left column) and $z_{REC}$ (right column) for Case {\it 1}. The first, second and third rows correspond to the production of hadrons with charge $-1$, $+1$ and $0$, respectively. Meanwhile, for $z\in(0.1,0.7)$, we note that the K-factor for $d\sigma^K/d\sigma^\pi$ tends to 1, which indicates that the higher-order effects are due to corrections in the partonic cross-sections and that the cuts imposed are eliminating an important part of the real-radiation contribution. As already mentioned, $z_{REC}$ has a smoother behaviour and corresponds to a physical observable, so it gives a more reliable phenomenological description.

Similarly, in Fig. \ref{fig:Figura4B}, we show the differential cross-section distributions w.r.t the momentum fraction $z$ imposing the restrictions associated to Case {\it 2}. However, this time the differences between the LO and NLO distributions are more noticeable: this is expected since we are looking at events that are being produced far from the Born-level kinematics. In particular, the NLO corrections are ${\cal O}(15 \, \%)$ larger than the LO ones for the $z_{REAL}$-spectrum associated to negative hadron production. In any case, we notice that this higher-order effects tend to reduce the $d\sigma^K/d \sigma^\pi$ in the low-$z$, but behave in the opposite direction for medium and high values of $z$. Also, for these kinematical cuts, there is an important difference between negative (first row) and positive (second row) hadron production, specially when considering the NLO contributions. Regarding the differences between $z_{REAL}$ and $z_{REC}$ spectra, the behaviour is very similar to the one observed for Case {\it 1}, pointing towards a reduced sensitivity in the kinematical cuts.

To conclude this Section, let us comment on the differences between Cases {\it 1} and {\it 2}. We have to keep in mind that our purpose is to disentangle the different partonic channels as much as possible, because we want to extract information about the FFs (and they depend on the flavour of the parton undergoing the hadronization process). We have seen that Case {\it 1} leads to an enhancement of $qg$-channel contribution compared to the default configuration (as well as w.r.t. Case {\it 2}). Since $qg$ channel is associated to the sub-processes $qg \to q + \gamma$ at LO and $qg \to q + g + \gamma$ at NLO, there is a high probability that the final state hadron comes from the hadronization of quark $q$. On the other hand, it is worth analyzing the relations between the $d\sigma^K/d \sigma^\pi$ ratios in both scenarios. In Fig. \ref{fig:Figura4C}, we study the ratio $R^{(1)}/R^{(2)}$, with
\beq 
R^{(1)} = {\left. \frac{d\sigma^K}{d\sigma^\pi} \right|} _{|\eta|<0.5} \ , \quad \quad R^{(2)} = {\left. \frac{d\sigma^K}{d\sigma^\pi} \right|} _{1.5<|\eta|<2} \ ,
\label{eq:RatiosRi}
\eeq
for LHC energies and different charges of the produced hadron. The red (blue) points represent the LO (NLO QCD + LO QED) predictions, whilst the dashed (solid) black lines corresponds to the respective linear trend of $R^{(1)}/R^{(2)}$ as a function of $z_{REC}$. We observe that the NLO QCD + LO QED corrections tend to decrease (increase) the cross-section for negative (positive) hadron production, and still the ratios are $\mathcal{O}(1.1)$ in all the cases. Since this effect is charge-dependent, it is originated by the $z$-dependence of the FFs and confirms the trend observed in Figs. \ref{fig:Figura4A}-\ref{fig:Figura4B}. Also, for neutral hadron production, we notice that LO and NLO QCD + LO QED corrections tends to cancel, preserving the same trend as a function of $z_{REC}$.


\section{Imposing constraints on FFs through cross-section ratios}
\label{sec:FFconstraints}
At this point, it is worth recalling the definition of the hadronic cross-section through the factorization theorem. In the particular case studied in this work, we have
\beqn 
\nn d\sigma^h  &=& \sum_{a_1,a_2,a_3} \int \, dx_1 dx_2 dz \, f^{H_1}_{a_1}(x_1) \, f^{H_2}_{a_2}(x_2)\, d^{h}_{a_3}(z) \,
\\ &\times& d\hat{\sigma}_{a_1 a_2\to a_3 \gamma}(x_1 P_1,x_2 P_2, P_h/z, P_{\gamma}) \, ,
\label{eq:XStotal}
\eeqn 
where $\{P_1,P_2\}$ are the momenta of the colliding hadrons $H_1$ and $H_2$, $P_\gamma$ is the momentum of the photon and $P_h$ is the momentum of the produced hadron and the hadronic center-of-mass energy is given by $E_{CM}=\sqrt{2 \, P_1 \cdot P_2}$. All the partonic dependence, including the phase-space integrals, the higher-order terms and the corresponding
measure functions\footnote{For a more detailed explanation of the inclusion of higher-order corrections, as well as the explicit formulae, we refer the interested reader to Refs. \cite{deFlorian:2010vy,Renteria-Estrada:2021rqp,Renteria-Estrada:2022scipost}.}, is embodied within $d\hat{\sigma}_{a_1 a_2\to a_3 \gamma}$. To keep the discussion simpler, we avoid writing explicitly the factorization scale dependence inside the PDFs and FFs. Then, let us consider the $z$-distributions for two different hadrons, $h_1$ and $h_2$. According to Eq. (\ref{eq:XStotal}),
\beqn 
\nn && \frac{d\sigma^{h_i}}{dz}  = \sum_{a_3} d^{h_i}_{a_3}(z) \, 
\\ \nn &&\times \left[\sum_{a_1,a_2} \int \, dx_1 dx_2 \, f^{H_1}_{a_1}(x_1) \, f^{H_2}_{a_2}(x_2)\,  d\hat{\sigma}_{a_1 a_2\to a_3 \gamma}\right] 
\\ && = \sum_{a_3} d^{h_i}_{a_3}(z) \, \times g_{a_3}(z) \, ,
\label{eq:XStotal2}
\eeqn 
where we define a new function, $g$, which is independent on the hadron produced in the final state. Having in mind Eq. (\ref{eq:XStotal2}) we can make two assumptions:
\begin{enumerate}
    \item $z=z_{REAL}$ has a rather similar behaviour compared to $z_{REC}$, and
    \item there is only one dominant partonic contribution. 
\end{enumerate}
By virtue of the first one, we can make the replacement $z \to z_{REC}$ without introducing relevant effects and keeping the factorization. This is motivated by the fact that $z_{REC}$ is strongly correlated with $z_{REAL}$, in particular in some specific kinematical regimes \cite{Renteria-Estrada:2022scipost}. Regarding the second assumption, we recall the discussion in Sec. \ref{ssec:Subchannels}. The point is that imposing proper cuts, we can enhance or suppress some partonic channels. In the case studied before, we realize that it is possible to enhance $qg$-channel, making it more than 10 times larger than the others. On top of that, we can make another approximation: the $u$-channel is dominant. This is not only supported by the fact that \emph{the protons contain more up than down quarks} (very naively speaking), but also because
\beq 
|{\cal M}_{ug\to u\gamma}|^2 = 4 \, |{\cal M}_{dg\to d\gamma}|^2 \, ,
\eeq
which is a consequence of producing a photon. With all of these, we can primarily settle in Case {\it 1} kinematics (where the $qg$-channel was enhanced), and claim that
\beq
R^{K/\pi}(d\sigma)=\frac{d\sigma^{K}/dz_{REC}}{d\sigma^{\pi}/dz_{REC}} \approx \frac{d^{K}_{u}(z_{REC})}{d^{\pi}_{u}(z_{REC})} = R^{K/\pi}(d_u)\, ,
\label{eq:MasterEQ}
\eeq
is a reasonably good approximation. During the rest of this Section, we will explore the validity of Eq. (\ref{eq:MasterEQ}). For the sake of simplicity, we will also restrict the higher-order corrections to NLO QCD. This is because the considered FFs do not include higher-order QED effects. Furthermore, it was found that their impact in the cross-section $pp \to \gamma + h$ was far below ${\cal O}(1 \, \%)$. So, it is safe to test Eq. (\ref{eq:MasterEQ}) up to NLO QCD accuracy.

Finally, we would like to emphasize that the assumptions considered here are independent of the nature of $h_1$ and $h_2$. In other words, we are comparing any pair of hadrons, regardless of their electromagnetic charge, isospin or composition. We expect deviations in the validity of Eq. (\ref{eq:MasterEQ}) originated in a very different composition of $h_1$ and $h_2$, and this is why we decided to compare same-sign hadrons. The presence of different valence quarks will result in a different weigth of other initial-state parton flavours, but still a noticeable enhancement of the $u$-started FFs is expected. This is the reason that explains the similarity of $\gamma + \pi^\pm$ and $\gamma + K^\pm$, and the slightly larger discrepancies between $\gamma + \pi^0$ and $\gamma + K^0$.


\begin{figure}[t!]
    \centering
    \includegraphics[width=0.49\textwidth]{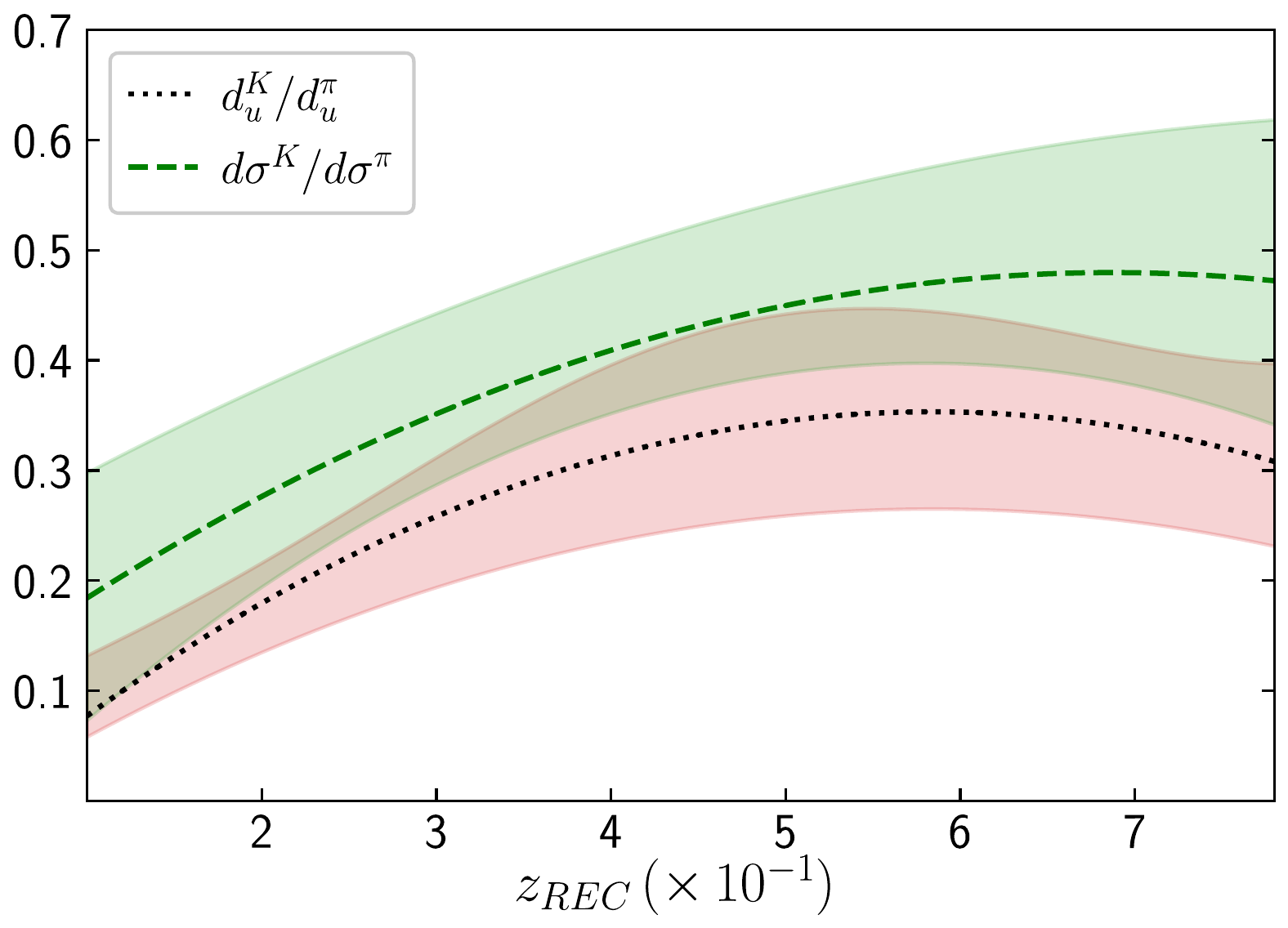}
    \includegraphics[width=0.49\textwidth]{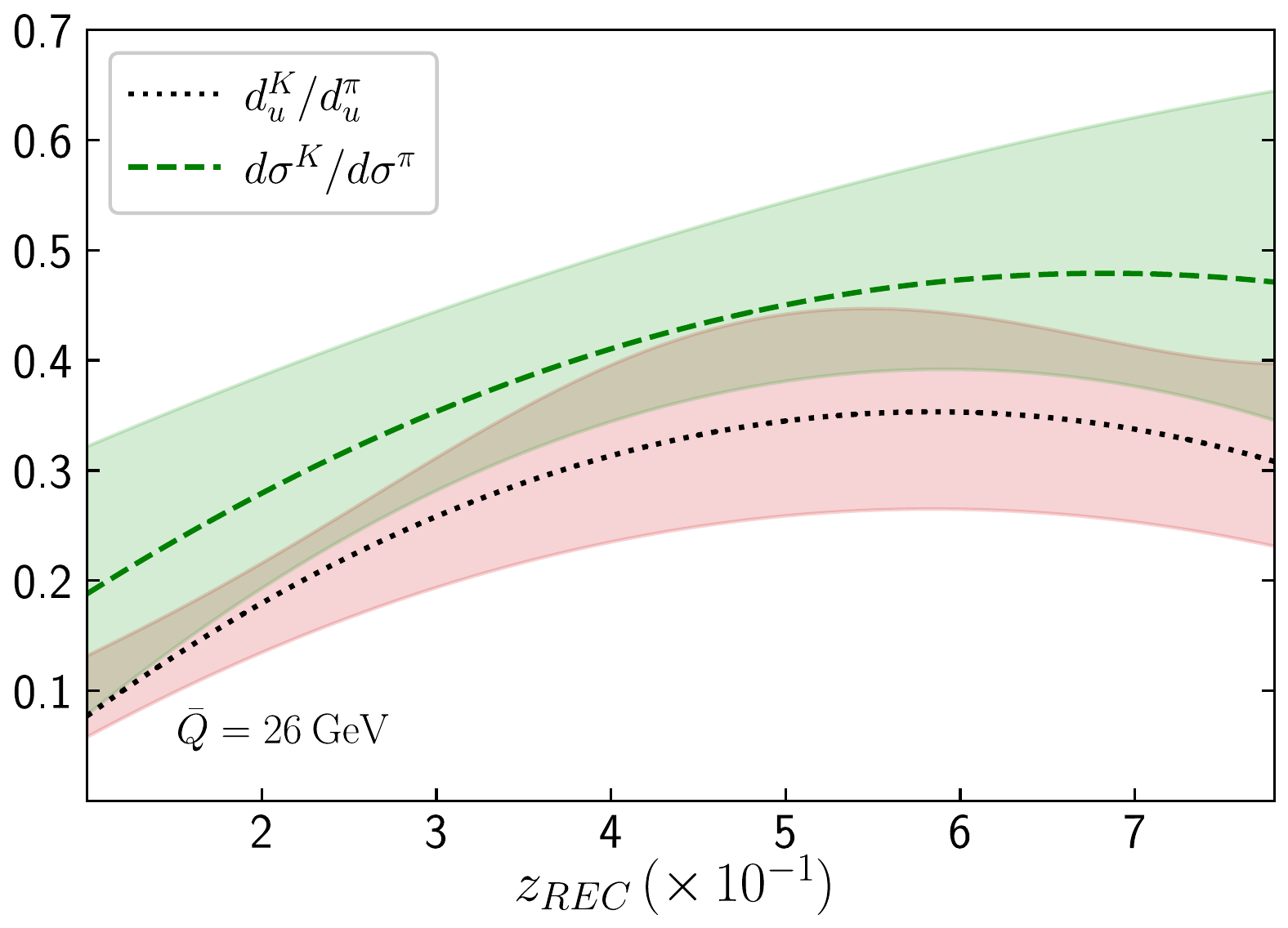}
    \caption{Comparison of $u$-started FF (dashed black line) and cross-section ratios (green dashed line) as a function of $z_{REC}$, for positive kaon and pion production up to NLO QCD accuracy. For the cross-section calculation we consider two scenarios: {\it 1)} with the default energy scale (upper plot), and {\it 2)} fixing the reference scale to $\bar{Q}=26\,{\rm GeV}$ (lower plot). The error bands are calculated as described in the main text.}
    \label{fig:FiguraFFxsnofixed}
\end{figure}

\subsection{Cross-section vs. FFs ratios}
\label{ssec:Comparison}
The first step in testing our approximations consists in the calculation of the FFs ratio (r.h.s. of Eq. (\ref{eq:MasterEQ})). Since FFs are sensitive to the energy scale, it is important to choose a reasonable reference value. Furthermore, this reference energy value must be related, in some sense, to the physical cross-sections. As we explained in Sec. \ref{sec:Analysis}, the default scale choice $\mu$ is the average of the transverse momenta of the produced hadron and the photon. Following Eq. (\ref{eq:MUdef}), this quantity depends on each individual event. Even if this situation is fully suitable for cross-section calculations, it is not the case for FFs determination. In fact, FFs fits start from a fixed-scale and then evolved through DGLAP equations\cite{Altarelli:1977zs,Dokshitzer:1977sg,Gribov:1972ri}. So, we can fix this issue by defining the energy scale average, $\bar{Q}$, as
\begin{eqnarray}
    \bar{Q} = \frac{\sum_i\, \mu(p_T^\gamma,p_T^h)_i \, (\sigma_{{\rm BIN}})_i}{\sigma_{{\rm TOTAL}}}\,,
\end{eqnarray}
where we are using information extracted from the histograms generated by the MC code (run with the default $\mu$ scale). For all the configurations considered in the previous Sections, we have $\bar{Q}=26\,{\rm GeV}$. In this way, we are fixing an energy scale for the MC that still contains physical information, and have well-defined FFs evaluated at the energy $\bar{Q}$. 

In Fig. \ref{fig:FiguraFFxsnofixed}, we present a comparison between the FFs and the cross-section ratios as a function of $z_{REC}$, for positive hadron production. As motivated in the previous discussion, we restrict our attention to the up-initiated fragmentation processes. Thus, we plot the ratio of the FFs $R^{K/\pi}(d_u)$ (dashed black lines) evaluated at the energy scale $\bar{Q}=26\,{\rm GeV}$, including up to NLO QCD corrections. In the same figure, we present the ratio of the cross-sections $R^{K/\pi}(d\sigma)$ up to NLO QCD accuracy (dashed green lines) in two different scenarios: {\it 1)} using the reference energy scale defined in Eq. (\ref{eq:MUdef}), i.e. the average transverse momenta of the produced particles (upper plot), and {\it 2)} fixing the reference energy scale as $\mu \equiv \bar{Q}$ (lower plot). The theoretical uncertainties associated to the cross-section ratios (green band) was calculated varying the energy scale by a factor two up and down, namely the standard procedure followed in Refs.~\cite{deFlorian:2010vy,Renteria-Estrada:2021rqp,Renteria-Estrada:2022scipost}. Meanwhile, the uncertainty for the ratio $R^{K/\pi}(d_u)$ (pink band) was estimated based on Ref.~\cite{deFlorian:2007aj}, where the reported errors were 10\% and 15\% for the pion and kaon fragmentation functions, respectively.

For the two scenarios considered in Fig. \ref{fig:FiguraFFxsnofixed}, the FF ratio oscillated in the range 0.1$-$0.35, reaching a maximum for $z_{REC} \approx 0.6$. In Scenario {\it 1)} (upper plot), the cross-section ratio closely follows the FF-ratio for $z_{REC} \leq 0.55$, although then $R^{K/\pi}(d\sigma)$ almost stabilizes around $\approx 0.48$ and $R^{K/\pi}(d_u)$ diminishes to $\approx 0.48$. The behaviour is rather similar for Scenario {\it 2)} (lower plot), being the ratio of the cross-sections slightly smaller in the high-$z$ region. The most appreciable difference between both scenarios comes from the error band: the band for Scenario {\it 2)} is wider than the one for Scenario {\it 1)} due to the fact that in the former we fixed the reference energy scale to $\bar{Q}$. It is important to notice that the error bands associated to both ratios overlap, indicating that they represent compatible quantities. In other words, this means that the ratio of the cross-sections is directly related to the ratio $R^{K/\pi}(d_u)$, and could be used to constrain the shape of these FFs. 

\begin{figure}[t!]
    \centering    
    \includegraphics[width=0.49\textwidth]{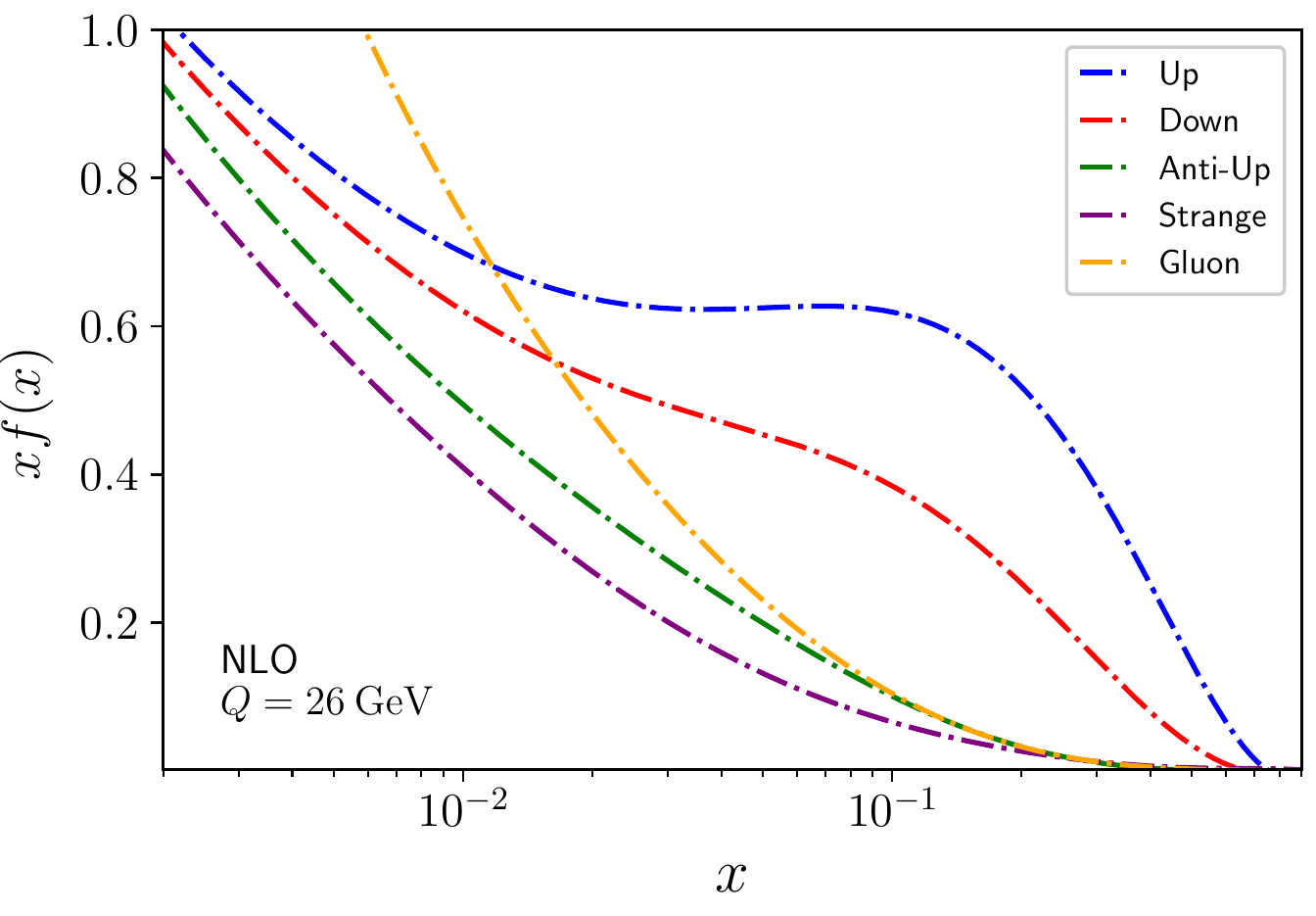} 
    \caption{Parton density functions (PDF) from the {\tt NNPDF31\_nlo\_as\_0118\_luxqed} set, evaluated at the scale energy $\bar{Q}=26\,{\rm GeV}$. We consider only the up, down, anti-up and strange flavours, which are the quark flavours contributing the most to the total cross-section of this process, as well as the gluon PDFs (divided by a factor 10 to match the scale of the other distributions).}
    \label{fig:Figura8}
\end{figure}

\begin{figure}[t!]
    \centering
    \includegraphics[width=0.49\textwidth]{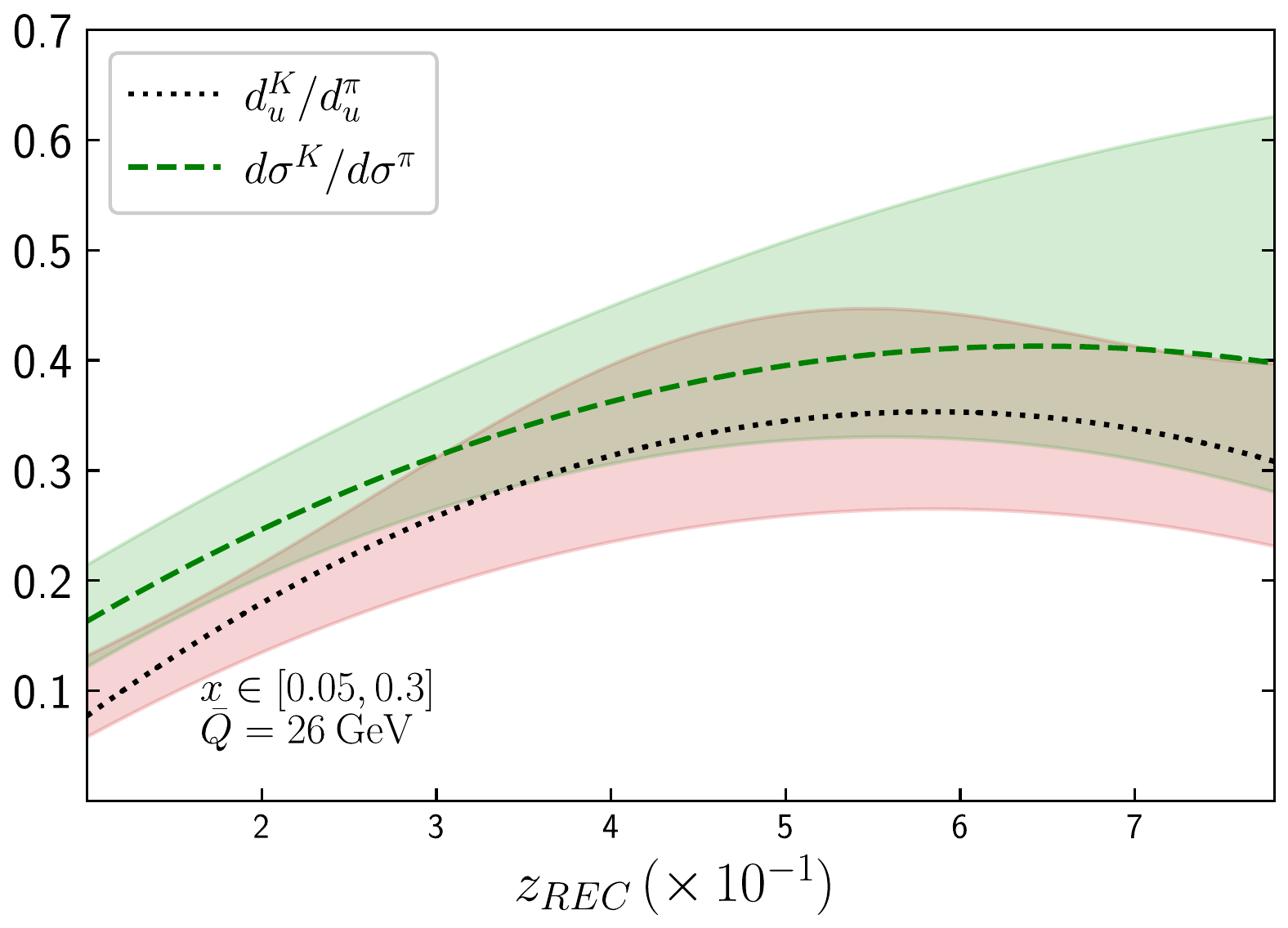}
    \caption{Comparison of $u$-started FF (dashed black line) and cross-section ratios (green dashed line) as a function of $z_{REC}$, for positive kaon and pion production up to NLO QCD accuracy. For the cross-section calculation we fixed the reference scale to $\bar{Q}=26\, {\rm GeV}$ and impose the restriction $x \in (0.05,0.3)$ on both colliding partons to enhance the contribution of the $u$-channel. The error bands are calculated as described in the main text.
    }
    \label{fig:FiguraFFxsIMPROVED}
\end{figure}

\begin{figure}[h!]
    \centering
    \includegraphics[width=0.49\textwidth]{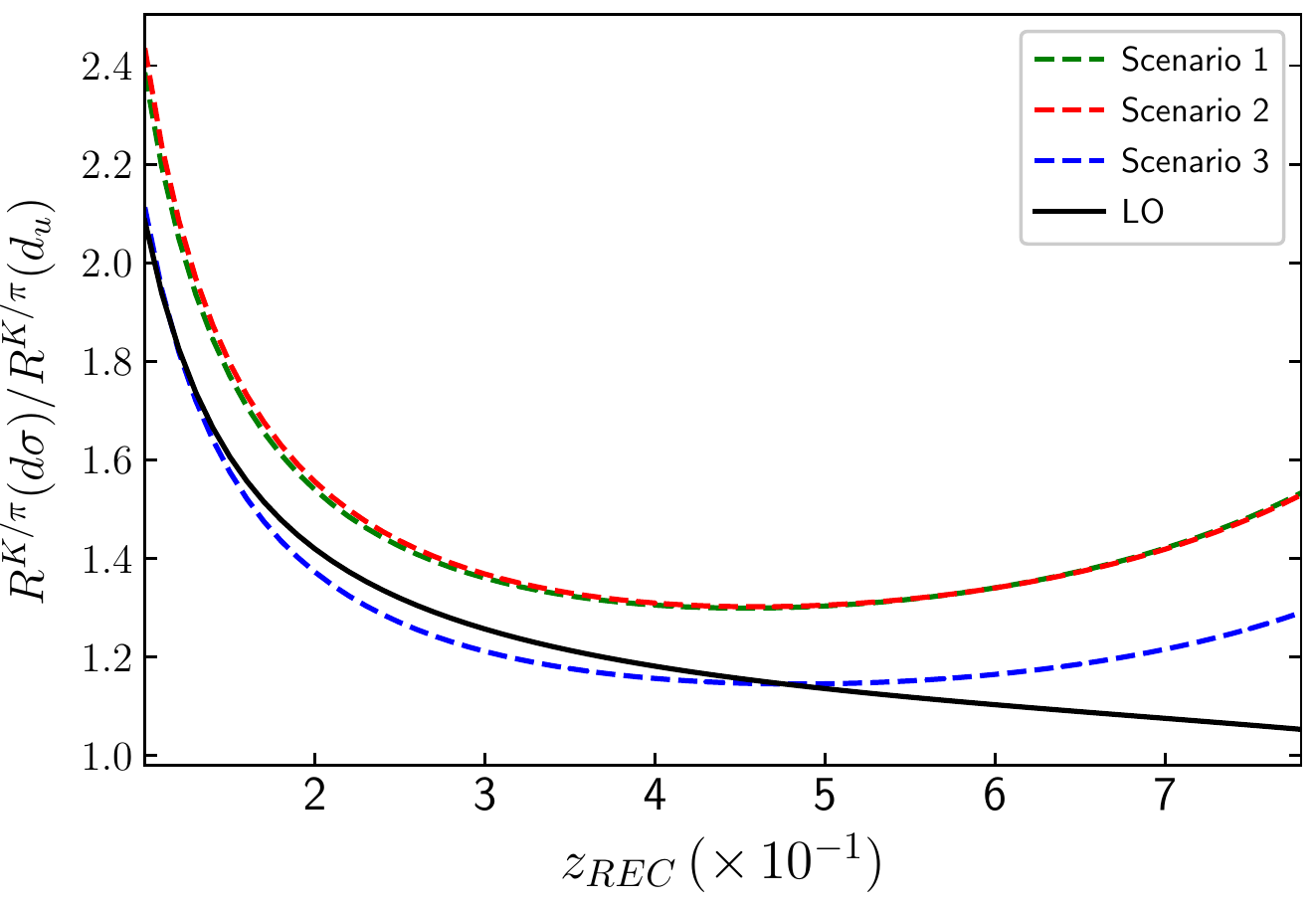}
    \caption{Ratio $R^{K/\pi}(d\sigma)/R^{K/\pi}(d_u)$, as a function of $z_{REC}$ for different Scenarios: \emph{1)} (green), \emph{2)} (red) and the physically-improved \emph{3)} (blue). We also include the LO ratio (black), with the configuration of Scenario \emph{3)}, to have an estimation of the impact of higher-order corrections.
    }
    \label{fig:FiguraCSratioFFratio}
\end{figure}


\subsection{Improved analysis with physically-motivated cuts}
\label{ssec:FFimproved}
In the previous two Scenarios, we observed that $R^{K/\pi}(d_u)$ and $R^{K/\pi}(d\sigma)$ have a very similar shape, with overlapping error bands. Still, the central values present a small offset, due to the contributions of different partonic channels, both during the collision (PDFs) as well during the hadronization process (FFs). So, we can give a step further and enhance the dominance of $u$-started processes by looking into the PDF shapes. For this, we consider the distributions of the {\tt NNPDF31\_nlo\_as\_0118\_luxqed} set in Fig. \ref{fig:Figura8}, where it is possible to appreciate that $u$ and $d$ PDFs differ the most in the region $x \in (0.05,0.3)$. In fact, in that range $f_u$ is roughly 50$\, \%$ larger than $f_d$, and almost 3 times larger than $f_{\bar u}$. The other flavours are even more suppressed w.r.t. $u$ PDF, and $f_g$ is roughly 10 times larger than any quark PDF. This means that, by restricting to the aforementioned region, the luminosity of the $ug$ channel is at least 50$\, \%$ bigger than $dg$ channel, and more than one order of magnitude larger w.r.t. the remaining contributions. 

For this reason, we decided to implement a cut and restrict the $x$ range. Since $x$ is not a physical variable, we relied on the discussion presented in Refs. \cite{deFlorian:2010vy,Renteria-Estrada:2022scipost} and used an approximation in terms of experimentally-accessible quantities. Explicitly, we have the reconstructed $x$ momentum fractions given by
\beqn
(x_1)_{REC} &=& p_T^{\gamma} \, \frac{\exp(\eta^{\pi}) + \exp(\eta^{\gamma})}{\sqrt{S_{CM}}} \, ,
\\ (x_2)_{REC} &=& p_T^{\gamma} \, \frac{\exp(-\eta^{\pi}) + \exp(-\eta^{\gamma})}{\sqrt{S_{CM}}} \, ,
\eeqn
which are strongly correlated with the real MC partonic-momentum fractions in a wide kinematical range \cite{Renteria-Estrada:2022scipost}. Then, we defined the Scenario \emph{3)} by imposing the cuts
\beq
0.03 \leq \{(x_1)_{REC},(x_2)_{REC}\} \leq 0.5 \, ,
\eeq
on top of the ones already applied for Scenario \emph{2)}. The results are presented in Fig. \ref{fig:FiguraFFxsIMPROVED}, with the corresponding error bands calculated following the procedure previously described. We immediately notice that $R^{K/\pi}(d_u)$ and $R^{K/\pi}(d\sigma)$ are much closer than in the previous scenarios, particularly in the range $z \in (0.35,0.65)$. Again, both ratios exhibit a rather similar shape, although a deviation is present in the low- and high-$z$ region. In any case, the error bands overlap for the whole $z$ range considered. This indicates that more stringent constraints can be imposed to $R^{K/\pi}(d_u)$ from the ratio of the cross-sections, paving the road for a more precise determination of heavy-meson FFs from experimental data.

Finally, we compared the three proposed scenarios among them, and w.r.t. the LO estimation within Scenario \emph{3)}. As we already explained, the $qg$-channel is the dominant one for photon-hadron production at colliders, being $qQ$ almost one order of magnitude smaller. At LO, the $qg$-channel is even more relevant because the other available channel is $q\bar{q}$, whose luminosity is highly suppressed by the presence of sea distributions. In Fig. \ref{fig:FiguraCSratioFFratio}, we present the ratio $R^{K/\pi}(d\sigma)/R^{K/\pi}(d_u)$ for Scenarios \emph{1)} (red), \emph{2)} (green), \emph{3)} (blue) and the LO estimation in Scenario \emph{3)} (black). If only $ug$-channel at LO contributes, then the ratio $R^{K/\pi}(d\sigma)/R^{K/\pi}(d_u)$ should be equal to 1, in agreement with Eq. (\ref{eq:MasterEQ}). Since the cuts associated to Scenario \emph{3)} enhance $ug$-channel, the complete LO contribution tends to 1 for high-$z$ values, in spite of deviating towards smaller values of $z$. It is important to notice that the NLO QCD corrections to the $R^{K/\pi}(d\sigma)/R^{K/\pi}(d_u)$ are also close to 1, showing an almost flat behaviour in the range $z \in (0.3,0.65)$. This, again, means that the FFs for kaons could be directly obtained by means of a constant re-scaling factor from the pion FFs, which is much more precisely determine from the experiments.


\section{Conclusions and outlook}
\label{sec:conclusions}
In this article, we analyzed strategies for imposing constrains on heavy-mesons FFs, based on our knowledge of hadron-photon production at colliders. We relied on the fact that parton-momentum fractions can be accurately described in terms of functions of experimentally-measurable quantities, as we demonstrated in Refs. \cite{deFlorian:2010vy,Renteria-Estrada:2022scipost}. 

Then, we center into the study of the $z_{REC}$ spectrum of $\gamma+h$ production at colliders, including up to NLO QCD + LO QED effects. We consider the cases $h=\{\pi,K\}$ since they are the lightest mesons, which involve larger cross-sections and smaller statistical errors. We studied the ratio of production rates for $\gamma + K$ w.r.t. $\gamma + \pi$, as a function of $z_{REC}$, varying the charge of the hadron and other kinematical cuts. We concluded that positive and negative hadron production have a different behaviour when NLO QCD corrections are taken into account. Also, we analyzed the contributions of the different partonic channels, identifying $qg$ as the dominant one.

With all this information, we proceed to study the relation between the kaon-to-pion FFs and cross-section rates, as a function of $z_{REC}$. First, we realize that $ug$-channel is favoured both by the luminosity as well as for the electromagnetic charge of $u$ w.r.t. $d$ quarks. Then, we defined three scenarios with different kinematical configurations and analysed the relations between the ratios $R^{K/\pi}(d\sigma)=d\sigma^{K}/d\sigma^{\pi} (z_{REC})$ vs. $R^{K/\pi}(d_u)=d^{K}_{u}(z_{REC})/d^{\pi}_{u}(z_{REC})$. By imposing a cut in the rapidity to concentrate events around the Born-like kinematics and requiring $0.03 \leq \{(x_1)_{REC},(x_2)_{REC}\} \leq 0.5$, we succeeded in enhancing even more the contribution of the $ug$-channel and isolate the $u$-initiated FFs. In fact, the results shown in Fig. \ref{fig:FiguraFFxsIMPROVED} suggest that the ratio of FFs is strongly constrained by the corresponding cross-section ratios. In other words, that it is possible to relate pion and kaon FFs by computing $R^{K/\pi}(d\sigma)$.

This works suggests that by using appropriate cuts it is possible to impose stringent constraints on the FFs. Here we present a concrete example of such constraints for $u$-initiated FFs, supporting the validity of the proposed strategy. It is worth highlighting that the proposed methodology is largely independent of the hadrons being compared, since it relies on the enhancement of certain production channels. Furthermore, we foresight that the application of ML-inspired selection cuts could help us to isolate contributions of other production and/or hadronization channels, thus allowing to determine with much more precision the FFs, shedding light into the underlying phenomena driving the hadronization process.

\section*{Acknowledgements}
We would like to thank L. Cieri, G. Rodrigo, R. Sassot and P. Zurita for fruitful comments about the preliminar version of this article.
The work of S.~A.~Ochoa-Oregon, D.~F.~Rentería-Estrada and R.~J.~Hern\'andez-Pinto is supported by CONACyT (Mexico) Project No. 320856 (Paradigmas y Controversias de la Ciencia 2022), Ciencia de Frontera 2021-2042 and by PROFAPI 2022 Grant No. PRO\_A1\_024 (Universidad Aut\'onoma de Sinaloa). R.~J.~Hern\'andez-Pinto is as well supported by Sistema Nacional de Investigadores of CONACyT (Mexico). The work of G.~Sborlini is partially supported by EU Horizon 2020 research and innovation program STRONG-2020 project under grant agreement No. 824093 and H2020-MSCA-COFUND-2020 USAL4EXCELLENCE-PROOPI-391 project under grant agreement No 101034371.

\bibliography{refs}
\end{document}